\documentclass[%
  reprint,
  superscriptaddress,
  showkeys,
  amsmath,amssymb,
  aps,pra,
  longbibliography,
  floatfix,
  x11names
]{revtex4-2}

\usepackage{xcolor}

\usepackage{graphicx}

\usepackage{tikz}
\usetikzlibrary{calc} 
\usetikzlibrary{intersections} 
\usetikzlibrary{decorations.markings}

\usepackage{orcidlink}
\usepackage{hyperref}
\hypersetup{hidelinks}

\newcommand{\ii}{\mathrm{i}}
\newtheorem{definition}{Definition}
\newtheorem{proposition}{Proposition}
\newenvironment{proof}{\par\noindent\textit{Proof.}\ }{\hfill$\square$\par}

\begin{document}

\title{Construction of Kochen--Specker Sets from Mutually Unbiased Bases}

\author{Mirko Navara\,\orcidlink{0000-0002-0880-5992}}
\email{navara@fel.cvut.cz}
\homepage{https://cmp.felk.cvut.cz/~navara}

\affiliation{Faculty of Electrical Engineering,
Czech Technical University in Prague,
Technick\'a 2,
CZ-166~27 Prague 6,
Czech Republic}

\author{Karl Svozil\,\orcidlink{0000-0001-6554-2802}}
\email{karl.svozil@tuwien.ac.at}
\homepage{http://tph.tuwien.ac.at/~svozil}

\affiliation{Institute for Theoretical Physics,
TU Wien,
Wiedner Hauptstra{\ss}e 8-10/136,
1040 Vienna,  Austria}

\begin{abstract}
We give a constructive analysis of Kochen--Specker (KS) contextuality at the level of complete orthogonal bases (contexts). In dimension three, three noncomputational mutually unbiased bases generate a union of $165$ rays and $130$ contexts. Exact provenance accounting gives a symmetric decomposition: the three $69$-ray lineages share $21$ rays and $10$ contexts, while each contributes $48$ exclusive rays and $40$ lineage-specific contexts. Each $69$-ray, $50$-context lineage is a KS configuration; as a hypergraph it is the $69$-$50$ KS set reported earlier by Pavi\v{c}i\'{c} and Megill. This framework clarifies the relation among the Yu--Oh, Harding--Salinas Schmeis, Pavi\v{c}i\'{c}--Megill, and Cabello constructions. In dimensions four and five, we give parametrized orthogonality gadgets whose connector contexts force a central ray to be maximally unbiased relative to the computational basis. For the concrete four-dimensional example, the $20$-ray gadget and the Cabello $18$-ray set are informationally equivalent subsets of the $24$-ray Peres--Mermin eigensystem, although their context hypergraphs have different two-valued-state behavior. The results show why both the rays and the complete incidence structure of their contexts are needed in assessments of KS contextuality.
\end{abstract}


\maketitle

\section{Introduction}

The Kochen--Specker (KS) theorem shows that certain finite families of quantum propositions do not admit a~context-independent assignment of classical truth values compatible with their orthogonality relations~\cite{kochen1}.  A~finite KS proof is therefore specified not only by a set of rays but also by the complete bases in which those rays occur.

\begin{definition}[Rays and contexts]
\label{def:contexts}
A \emph{ray} is a one-dimensional subspace
$[v]=\{\lambda v:\lambda\in\mathbb{C}\}$ generated by a
nonzero representative \ensuremath{v\in\mathbb{C}^D\setminus\{0\}}.
The scalar zero is included because the ray is a linear subspace.
A~\emph{context} (block) in $\mathbb{C}^D$ is a set of $D$ mutually orthogonal rays spanning the space.  In the associated $D$-uniform hypergraph the rays are vertices and the contexts are hyperedges.  A ray is \emph{intertwining} if it belongs to at least two contexts; a ray occurring in exactly one context is nonintertwining.
\end{definition}

\begin{definition}[Two-valued states]
\label{def:states}
For a context hypergraph $(V,E)$, a two-valued state is a map $s:V\to\{0,1\}$ satisfying $\sum_{v\in S}s(v)=1$ for every \ensuremath{S\in E}.
A coordinatized hypergraph with no two-valued state is a KS configuration.  A family of two-valued states is \emph{separating} if every pair of distinct rays receives different values in at least one state; it is \emph{unital} if every ray receives the value $1$ in at least one state.
\end{definition}

These definitions distinguish the logical KS criterion from resource measures used in inequalities, chromatic formulations, or nonlocal games~\cite{cabello:210401,2024-cabello-trandafir0,svozil-2025-color}.  Reduced incidence structures can be useful in those settings, but deleting singly occurring rays or omitting available contexts can change the set of two-valued states.  For example, the seven-ray subconfiguration discussed in Ref.~\cite[Fig.~6(a)]{pavicic-2025} has different classical-state behavior from its thirteen-ray completion.  We therefore state explicitly which context hypergraph is being analyzed at each step.

The paper has two main contributions:
\begin{enumerate}
\item In $\mathbb{C}^3$, three noncomputational mutually unbiased bases (MUBs), together with a fixed family of algebraic generation maps, produce $225$ labelled vectors.  Projective identification reduces these to $165$ rays, and an exhaustive orthogonality search gives $130$ contexts.  The provenance decomposition is symmetric: $21$ rays and $10$ contexts are common to all three $69$-ray lineages, while each lineage has $48$ exclusive rays and $40$ lineage-specific contexts.  Each lineage contains $69$ rays and $50$ contexts and is uncolorable.  As a hypergraph, each lineage is the $69$-$50$ KS set reported earlier by Pavi\v{c}i\'{c} and Megill~\cite{pavicic-megill-2022,Pavicic_2023}.
\item In $D=4$ and $D=5$, connector contexts convert orthogonality into linear equations for $|x_j|^2$, forcing a central vector $u=(x_1,\ldots,x_D)$ to be maximally unbiased relative to the computational basis.  In the concrete $D=4$ case, the resulting $20$-ray set and the Cabello $18$-ray KS set are mutually reconstructible inside the $24$-ray Peres--Mermin eigensystem~\cite{peres111,mermin90b,peres-91,pavicic-2004ksafq,cabello-96}; nevertheless, their context hypergraphs have different two-valued-state spaces.
\end{enumerate}

Section~\ref{2025-yohscg} relates the
three-dimensional construction for a qutrit (a three-level quantum system)
to the Yu--Oh, Harding--Salinas Schmeis, and Cabello diagrams and states the generation rules.  Section~\ref{sec:inventory} gives the complete ray and context inventories and their provenance decomposition.  Section~\ref{2025-hdfg} separates the higher-dimensional forcing constraints from their coordinate completions and then compares the $D=4$ gadget with the Cabello set.  Appendix~\ref{appendix:MUBsD3} records the qutrit MUB convention used throughout.

\section{The Yu--Oh--Harding--Salinas Schmeis--Cabello gadget}
\label{2025-yohscg}

The $25$-ray, $16$-context full-context completion of the original $13$-ray Yu--Oh (YO) set supports a state-independent noncontextuality inequality for a qutrit~\cite{PhysRevLett.103.050401,Yu-2012,Yu2015}, but it is not itself a KS configuration under Definition~\ref{def:states} because it has $24$ two-valued states, which are separating and unital.  Cabello's ``triple YO diagram''~\cite{Cabello2025} combines three copies of this $25$-ray full-context completion whose center rays form the Fourier basis
\[
\mathcal{B}_1=\{(1,1,1),(1,\omega,\omega^2),(1,\omega^2,\omega)\},
\qquad \omega=\mathrm{e}^{2\pi\ii{}/3}.
\]
Relative to the computational rim $\mathcal{B}_0=\{(1,0,0),(0,1,0),(0,0,1)\}$, this is one member of a complete set of four MUBs.  Replacing $\mathcal{B}_1$ by either of the other two noncomputational MUBs listed in Appendix~\ref{appendix:MUBsD3} gives an isomorphic $69$-ray, $50$-context KS configuration.  This hypergraph was found earlier.  Pavi\v{c}i\'{c} and Megill listed a critical $69$-$50$ KS hypergraph as the smallest one generated by the vector components $\{0,\pm\omega,2\omega,\pm\omega^2,2\omega^2\}$~\cite[Table~I]{pavicic-megill-2022}, and Pavi\v{c}i\'{c} gave its diagram, its hypergraph string, and a complex coordinatization~\cite[Fig.~10(e), Table~2, and Appendix~C]{Pavicic_2023}.  We have checked that each of our lineages is isomorphic to that $69$-$50$ hypergraph.  Deleting its $36$ nonintertwining rays leaves the $33$ rays and $14$ complete contexts of the set in Ref.~\cite{Cabello2025}; see also Refs.~\cite{Pavicic-2025-simplest,Pavicic-2026-entropy}.  The same incidence pattern appears in the Harding--Salinas Schmeis construction~\cite[Fig.~2 and Theorem~2.6]{harding2025remarksIJTP}; Fig.~\ref{fig:yohs} uses their Greechie-like notation to display the common structure.

\paragraph*{Guide to the construction.}
The calculation has four stages.  
First, index the nine seed vectors 
by $j=1,\ldots,9$, and denote 
the seed vector with index $j$ by 
$\sigma_j$, where
\[
\begin{aligned}
&\ensuremath{(\sigma_1,\sigma_2,\sigma_3)=((1,1,1),(1,\omega,\omega^2),(1,\omega^2,\omega))},\\
&\ensuremath{(\sigma_4,\sigma_5,\sigma_6)=((1,\omega,\omega),(1,\omega^2,1),(1,1,\omega^2))},\\
&\ensuremath{(\sigma_7,\sigma_8,\sigma_9)=((1,\omega^2,\omega^2),(1,\omega,1),(1,1,\omega))}.
\end{aligned}
\]
The three displayed triples are, respectively, the bases \(\mathcal B_1,\mathcal B_2,\mathcal B_3\).
Second, the $25$ generation rules given below and in Table~\ref{tab:vector_generation} are applied to every seed, producing $225$ labelled vectors.  Third, scalar multiples are identified projectively, leaving the $165$ rays in Table~\ref{tab:noncollinear_vectors_6col}; all mutually orthogonal triples among those rays are then enumerated, giving the $130$ contexts in Table~\ref{tab:orthogonal_bases}.  Fourth, every ray and context is classified by the complete set of seed lineages from which it can be generated.

This last step is essential.  The three $69$-ray lineages have a $21$-ray intersection, and no ray belongs to exactly two lineages.  Each lineage therefore consists of the same $21$ common rays plus $48$ exclusive rays.  At the context level, $10$ contexts are common to all three lineages and each lineage contributes $40$ additional contexts, so $10+3\cdot40=130$.  The provenance structure is therefore symmetric.

We work projectively throughout and use the Harding--Salinas Schmeis labels, augmented where needed by the symbols $e_i$.  The full tables make the computational claims reproducible, while the propositions below isolate the conclusions needed later.

\paragraph*{Generation rules.}
All rules use the Hermitian (complex) cross product~\cite{harding2025remarksIJTP}
\[
x\times y=\bigl(\overline{x_2y_3-x_3y_2},\,\overline{x_3y_1-x_1y_3},\,\overline{x_1y_2-x_2y_1}\bigr),
\]
which is orthogonal to both $x$ and $y$ with respect to $\langle x,y\rangle=\sum_m\overline{x_m}\,y_m$.  With $a_1,a_2,a_3$ the computational basis $\mathcal{B}_0$ and $u=\sigma_j$, the rules read ($i=1,2,3$ in the first row)
\begin{align*}
c_i&=-a_i\times u, & b_i&=a_i\times c_i, & d_i&=u\times c_i,\\
b_{12}&=b_1\times b_2, & b_{13}&=-b_1\times b_3, & b_{23}&=b_2\times b_3,\\
e_1&=c_1\times b_{23}, & e_2&=c_2\times b_{13}, & e_3&=c_3\times b_{12},\\
b_{112}&=-b_1\times b_{12}, & b_{113}&=b_1\times b_{13}, & b_{212}&=b_2\times b_{12},\\
b_{223}&=-b_2\times b_{23}, & b_{313}&=-b_3\times b_{13}, & b_{323}&=b_3\times b_{23}.
\end{align*}
Each new ray is orthogonal to its two inputs and shares a context with each of them in Fig.~\ref{fig:yohs}: $c_i$ is joined to $a_i$ and to $u$, and $b_{ij}$ to $b_i$ and to $b_j$, through two different contexts, whereas every other rule returns the third ray of a single context.  The signs only fix the representatives listed in Table~\ref{tab:vector_generation}; projectively they are immaterial.  Applied to the nine seeds, these rules reproduce all $225$ entries of Table~\ref{tab:vector_generation}.

\begin{figure}[ht]
\centering
\resizebox{0.40\textwidth}{!}{
\begin{tikzpicture}[ultra thick, scale=0.5]

    \colorlet{colorCircle}{gray}
    \colorlet{colorAC1}{red}
    \colorlet{colorCU1}{orange}
    \colorlet{colorAC2}{blue}
    \colorlet{colorCU2}{cyan}
    \colorlet{colorAC3}{green!60!black}
    \colorlet{colorCU3}{lime!70!black}
    \colorlet{colorB1B13}{magenta}
    \colorlet{colorB13B3}{violet}
    \colorlet{colorB3B23}{teal}
    \colorlet{colorB23B2}{olive}
    \colorlet{colorB2B12}{brown}
    \colorlet{colorB12B1}{pink}
    \colorlet{colorC3B12}{purple}
    \colorlet{colorC2B13}{black}   
    \colorlet{colorC1B23}{black!60}

    \def\R{10} \def\r{5} \def\rB{7.5} \def\rD{2.5} \def\shiftangle{45}
    \def\aOneAngle{90} \def\aTwoAngle{210} \def\aThreeAngle{330}

    \coordinate (u) at (0,0);
    \foreach \angle/\i in {\aOneAngle/1, \aTwoAngle/2, \aThreeAngle/3} {
        \coordinate (a\i) at (\angle:\R);
        \coordinate (c\i) at ({\angle+\shiftangle}:\r);
        \coordinate (d\i) at ({\angle+\shiftangle}:\rD);
    }
    \foreach \i in {1,2,3} {
        \coordinate (b\i) at ($(a\i)!0.5!(c\i)$);
    }
    \pgfmathsetmacro{\bOneAngle}{(\aOneAngle + (\aOneAngle+\shiftangle))/2}
    \pgfmathsetmacro{\bTwoAngle}{(\aTwoAngle + (\aTwoAngle+\shiftangle))/2}
    \pgfmathsetmacro{\bThreeAngle}{(\aThreeAngle + (\aThreeAngle+\shiftangle))/2}
    \pgfmathsetmacro{\bOneTwoAngle}{(\bOneAngle+\bTwoAngle)/2}
    \pgfmathsetmacro{\bTwoThreeAngle}{(\bTwoAngle+\bThreeAngle)/2}
    \pgfmathsetmacro{\bOneThreeAngle}{mod((\bThreeAngle + (\bOneAngle+360))/2, 360)}
    \coordinate (b12) at (\bOneTwoAngle:\rB);
    \coordinate (b23) at (\bTwoThreeAngle:\rB);
    \coordinate (b13) at (\bOneThreeAngle:\rB);
    \coordinate (b113) at ($(b1)!0.5!(b13)$); \coordinate (b313) at ($(b3)!0.5!(b13)$);
    \coordinate (b323) at ($(b3)!0.5!(b23)$); \coordinate (b223) at ($(b2)!0.5!(b23)$);
    \coordinate (b212) at ($(b2)!0.5!(b12)$); \coordinate (b112) at ($(b1)!0.5!(b12)$);

    \draw[colorCircle] (0,0) circle (\R);
    \draw[colorAC1] (a1) -- (c1); \draw[colorCU1] (c1) -- (u);
    \draw[colorAC2] (a2) -- (c2); \draw[colorCU2] (c2) -- (u);
    \draw[colorAC3] (a3) -- (c3); \draw[colorCU3] (c3) -- (u);
    \draw[colorB1B13] (b1) -- (b13); \draw[colorB13B3] (b13) -- (b3);
    \draw[colorB3B23] (b3) -- (b23); \draw[colorB23B2] (b23) -- (b2);
    \draw[colorB2B12] (b2) -- (b12); \draw[colorB12B1] (b12) -- (b1);

    \draw[colorC3B12] (c3) -- (b12);
    \draw[colorC2B13] (c2) -- (b13);
    \draw[colorC1B23] (c1) -- (b23);

    \node[circle, fill=colorCU1, inner sep=1.5pt, label={0:$d_1$}] at (d1) {};
    \node[circle, fill=colorCU2, inner sep=1.5pt, label={180:$d_2$}] at (d2) {};
    \node[circle, fill=colorCU3, inner sep=1.5pt, label={285:$d_3$}] at (d3) {};
    \node[circle, fill=colorB1B13, inner sep=1.5pt, label=above:$b_{113}$] at (b113) {};
    \node[circle, fill=colorB13B3, inner sep=1.5pt, label=right:$b_{313}$] at (b313) {};
    \node[circle, fill=colorB3B23, inner sep=1.5pt, label=right:$b_{323}$] at (b323) {};
    \node[circle, fill=colorB23B2, inner sep=1.5pt, label=below:$b_{223}$] at (b223) {};
    \node[circle, fill=colorB2B12, inner sep=1.5pt, label=right:$b_{212}$] at (b212) {};
    \node[circle, fill=colorB12B1, inner sep=1.5pt, label={[label distance=2mm]90:$b_{112}$}] at (b112) {};

    \node[circle, fill=colorCircle, inner sep=2.5pt, label={90:$a_1\equiv z_1$}] at (a1) {}; \node[circle, fill=colorAC1, inner sep=1.5pt] at (a1) {};
    \node[circle, fill=colorCircle, inner sep=2.5pt, label={210:$a_2\equiv z_2$}] at (a2) {}; \node[circle, fill=colorAC2, inner sep=1.5pt] at (a2) {};
    \node[circle, fill=colorCircle, inner sep=2.5pt, label={330:$a_3\equiv z_3$}] at (a3) {}; \node[circle, fill=colorAC3, inner sep=1.5pt] at (a3) {};

    \node[circle, fill=colorCU1, inner sep=4.5pt, label={[label distance=-1mm]275:$u\equiv h_0$}] at (u) {};
    \node[circle, fill=colorCU2, inner sep=3pt] at (u) {};
    \node[circle, fill=colorCU3, inner sep=1.5pt] at (u) {};
    \node[circle, fill=colorAC1, inner sep=3pt, label={[label distance=-1mm]110:$b_1\equiv y_+^1$}] at (b1) {}; \node[circle, fill=colorB1B13, inner sep=2pt] at (b1) {}; \node[circle, fill=colorB12B1, inner sep=1pt] at (b1) {};
    \node[circle, fill=colorAC2, inner sep=3pt, label={[label distance=-1mm]230:$b_2\equiv y_+^2$}] at (b2) {}; \node[circle, fill=colorB2B12, inner sep=2pt] at (b2) {}; \node[circle, fill=colorB23B2, inner sep=1pt] at (b2) {};
    \node[circle, fill=colorAC3, inner sep=3pt, label={[label distance=-1mm]0:$b_3\equiv y_+^3$}] at (b3) {}; \node[circle, fill=colorB3B23, inner sep=2pt] at (b3) {}; \node[circle, fill=colorB13B3, inner sep=1pt] at (b3) {};

    \node[circle, fill=colorAC1, inner sep=3pt, label={0:$c_1\equiv y_-^1$}] at (c1) {}; \node[circle, fill=colorCU1, inner sep=2pt] at (c1) {}; \node[circle, fill=colorC1B23, inner sep=1pt] at (c1) {};
    \node[circle, fill=colorAC2, inner sep=3pt, label={[label distance=-2mm]-5:$c_2\equiv y_-^2$}] at (c2) {}; \node[circle, fill=colorCU2, inner sep=2pt] at (c2) {}; \node[circle, fill=colorC2B13, inner sep=1pt] at (c2) {};
    \node[circle, fill=colorAC3, inner sep=3pt, label={[label distance=-2.5mm]93:$c_3\equiv y_-^3$}] at (c3) {}; \node[circle, fill=colorCU3, inner sep=2pt] at (c3) {}; \node[circle, fill=colorC3B12, inner sep=1pt] at (c3) {};

    \node[circle, fill=colorB12B1, inner sep=3pt, label={above left:$b_{12}$}] at (b12) {};
    \node[circle, fill=colorB12B1, inner sep=3pt, label={left:$\equiv h_3$}] at (b12) {};
    \node[circle, fill=colorB2B12, inner sep=2pt] at (b12) {}; \node[circle, fill=colorC3B12, inner sep=1pt] at (b12) {};
    \node[circle, fill=colorB23B2, inner sep=3pt, label=right:$b_{23}\equiv h_1$] at (b23) {}; \node[circle, fill=colorB3B23, inner sep=2pt] at (b23) {}; \node[circle, fill=colorC1B23, inner sep=1pt] at (b23) {};
    \node[circle, fill=colorB13B3, inner sep=3pt, label=right:$b_{13}\equiv h_2$] at (b13) {}; \node[circle, fill=colorB1B13, inner sep=2pt] at (b13) {}; \node[circle, fill=colorC2B13, inner sep=1pt] at (b13) {};

    \node[circle, fill=colorC1B23, inner sep=2.5pt, label={above right:$e_1$}] (e1) at ($(c1)!0.75!(b23)$) {};

    \node[circle, fill=colorC2B13, inner sep=2.5pt, label={left:$e_2$}] (e2) at ($(c2)!0.75!(b13)$) {};

    \node[circle, fill=colorC3B12, inner sep=2.5pt, label={below:$e_3$}] (e3) at ($(c3)!0.75!(b12)$) {};

\end{tikzpicture}
}
\caption{The $25$-ray full-context completion of the original $13$-ray Yu--Oh set~\cite{Yu-2012}, drawn in the Harding and Salinas Schmeis $3$-uniform hypergraph scheme~\cite{harding2025remarksIJTP}.}
\label{fig:yohs}
\end{figure}

\begin{table*}[htbp]
\caption{\label{tab:vector_generation}
\textit{Systematic generation of the vector set for contextuality analysis.}
This table details the complete set of $225$ vectors that form the basis of our structural analysis.
The $25$ generation rules \(F\) of Sec.~\ref{2025-yohscg} (rows) are applied to nine seed vectors \(\sigma_j\), indexed by \(j=1,\ldots,9\). Column \(j\) contains the vectors generated from the seed \(\sigma_j\). Columns 1--3, 4--6, and 7--9 correspond, respectively, to the three noncomputational MUBs
\(
\mathcal B_1=\{\sigma_1,\sigma_2,\sigma_3\},\
\mathcal B_2=\{\sigma_4,\sigma_5,\sigma_6\},\
\mathcal B_3=\{\sigma_7,\sigma_8,\sigma_9\}.
\)
Each basis generates a $69$-ray, $50$-context KS configuration. The entry in column \(j\) is \(F(\sigma_j)\) and receives the inventory label \(F_j\); for example, \(d_1(\sigma_4)\) is labelled \(d_{14}\). The identity rule \(u\equiv h_0\) gives \(u_j=\sigma_j\).
The final column counts the number of unique rays produced by each function,
highlighting the differing projective variety of the constructions.
The $165$ unique rays drawn from this set form the inventory used to identify orthogonal bases, hyperedges, or contexts.
The row symbols follow the Harding--Salinas Schmeis scheme~\cite{harding2025remarksIJTP},
with equivalences to the Yu--Oh numbering scheme~\cite{Yu-2012}.
}
\begin{ruledtabular}
\begin{tabular}{ccccccccccc}
seed&
      \multicolumn{3}{c|}{$\mathcal{B}_1$}&
      \multicolumn{3}{c|}{$\mathcal{B}_2$}&
      \multicolumn{3}{c}{$\mathcal{B}_3$}\\
\hline
rule & 
$j=1$ &
$j=2$ &
$j=3$ &
$j=4$ &
$j=5$ &
$j=6$ &
$j=7$ &
$j=8$ &
$j=9$ &
\#  \\ \hline
$a_1\equiv z_1$ & $(1,0,0)$ & $(1,0,0)$ & $(1,0,0)$ & $(1,0,0)$ & $(1,0,0)$ & $(1,0,0)$ & $(1,0,0)$ & $(1,0,0)$ & $(1,0,0)$ & 1 \\
$a_2\equiv z_2$ & $(0,1,0)$ & $(0,1,0)$ & $(0,1,0)$ & $(0,1,0)$ & $(0,1,0)$ & $(0,1,0)$ & $(0,1,0)$ & $(0,1,0)$ & $(0,1,0)$ & 1 \\
$a_3\equiv z_3$ & $(0,0,1)$ & $(0,0,1)$ & $(0,0,1)$ & $(0,0,1)$ & $(0,0,1)$ & $(0,0,1)$ & $(0,0,1)$ & $(0,0,1)$ & $(0,0,1)$ & 1 \\
$u\equiv h_0$ & $(1,1,1)$ & $(1,\omega,\omega^2)$ & $(1,\omega^2,\omega)$ & $(1,\omega,\omega)$ & $(1,\omega^2,1)$ & $(1,1,\omega^2)$ & $(1,\omega^2,\omega^2)$ & $(1,\omega,1)$ & $(1,1,\omega)$ & 9 \\
$b_1\equiv y_+^1$ & $(0,1,1)$ & $(0,\omega,\omega^2)$ & $(0,\omega^2,\omega)$ & $(0,\omega,\omega)$ & $(0,\omega^2,1)$ & $(0,1,\omega^2)$ & $(0,\omega^2,\omega^2)$ & $(0,\omega,1)$ & $(0,1,\omega)$ & 3 \\
$b_2\equiv y_+^2$ & $(1,0,1)$ & $(1,0,\omega^2)$ & $(1,0,\omega)$ & $(1,0,\omega)$ & $(1,0,1)$ & $(1,0,\omega^2)$ & $(1,0,\omega^2)$ & $(1,0,1)$ & $(1,0,\omega)$ & 3 \\
$b_3\equiv y_+^3$ & $(1,1,0)$ & $(1,\omega,0)$ & $(1,\omega^2,0)$ & $(1,\omega,0)$ & $(1,\omega^2,0)$ & $(1,1,0)$ & $(1,\omega^2,0)$ & $(1,\omega,0)$ & $(1,1,0)$ & 3 \\
$c_1\equiv y_-^1$ & $(0,1,-1)$ & $(0,\omega,-\omega^2)$ & $(0,\omega^2,-\omega)$ & $(0,\omega^2,-\omega^2)$ & $(0,1,-\omega)$ & $(0,\omega,-1)$ & $(0,\omega,-\omega)$ & $(0,1,-\omega^2)$ & $(0,\omega^2,-1)$ & 3 \\
$c_2\equiv y_-^2$ & $(-1,0,1)$ & $(-\omega,0,1)$ & $(-\omega^2,0,1)$ & $(-\omega^2,0,1)$ & $(-1,0,1)$ & $(-\omega,0,1)$ & $(-\omega,0,1)$ & $(-1,0,1)$ & $(-\omega^2,0,1)$ & 3 \\
$c_3\equiv y_-^3$ & $(1,-1,0)$ & $(\omega^2,-1,0)$ & $(\omega,-1,0)$ & $(\omega^2,-1,0)$ & $(\omega,-1,0)$ & $(1,-1,0)$ & $(\omega,-1,0)$ & $(\omega^2,-1,0)$ & $(1,-1,0)$ & 3 \\
$d_1$ & $(-2,1,1)$ & $(-2,\omega,\omega^2)$ & $(-2,\omega^2,\omega)$ & $(-2,\omega,\omega)$ & $(-2,\omega^2,1)$ & $(-2,1,\omega^2)$ & $(-2,\omega^2,\omega^2)$ & $(-2,\omega,1)$ & $(-2,1,\omega)$ & 9 \\
$d_2$ & $(1,-2,1)$ & $(\omega^2,-2,\omega)$ & $(\omega,-2,\omega^2)$ & $(\omega^2,-2,1)$ & $(\omega,-2,\omega)$ & $(1,-2,\omega^2)$ & $(\omega,-2,1)$ & $(\omega^2,-2,\omega^2)$ & $(1,-2,\omega)$ & 9 \\
$d_3$ & $(1,1,-2)$ & $(\omega,\omega^2,-2)$ & $(\omega^2,\omega,-2)$ & $(\omega^2,1,-2)$ & $(1,\omega^2,-2)$ & $(\omega,\omega,-2)$ & $(\omega,1,-2)$ & $(1,\omega,-2)$ & $(\omega^2,\omega^2,-2)$ & 9 \\
$b_{12}\equiv h_3$ & $(1,1,-1)$ & $(1,\omega,-\omega^2)$ & $(1,\omega^2,-\omega)$ & $(\omega,\omega^2,-\omega^2)$ & $(\omega,1,-\omega)$ & $(\omega,\omega,-1)$ & $(\omega^2,\omega,-\omega)$ & $(\omega^2,1,-\omega^2)$ & $(\omega^2,\omega^2,-1)$ & 9 \\
$b_{13}\equiv h_2$ & $(1,-1,1)$ & $(1,-\omega,\omega^2)$ & $(1,-\omega^2,\omega)$ & $(\omega,-\omega^2,\omega^2)$ & $(\omega,-1,\omega)$ & $(\omega,-\omega,1)$ & $(\omega^2,-\omega,\omega)$ & $(\omega^2,-1,\omega^2)$ & $(\omega^2,-\omega^2,1)$ & 9 \\
$b_{23}\equiv h_1$ & $(-1,1,1)$ & $(-1,\omega,\omega^2)$ & $(-1,\omega^2,\omega)$ & $(-\omega,\omega^2,\omega^2)$ & $(-\omega,1,\omega)$ & $(-\omega,\omega,1)$ & $(-\omega^2,\omega,\omega)$ & $(-\omega^2,1,\omega^2)$ & $(-\omega^2,\omega^2,1)$ & 9 \\
$e_1$ & $(2,1,1)$ & $(2,\omega,\omega^2)$ & $(2,\omega^2,\omega)$ & $(2\omega^2,1,1)$ & $(2\omega^2,\omega,\omega^2)$ & $(2\omega^2,\omega^2,\omega)$ & $(2\omega,1,1)$ & $(2\omega,\omega^2,\omega)$ & $(2\omega,\omega,\omega^2)$ & 9 \\
$e_2$ & $(1,2,1)$ & $(\omega^2,2,\omega)$ & $(\omega,2,\omega^2)$ & $(\omega,2\omega^2,\omega^2)$ & $(1,2\omega^2,1)$ & $(\omega^2,2\omega^2,\omega)$ & $(\omega^2,2\omega,\omega)$ & $(1,2\omega,1)$ & $(\omega,2\omega,\omega^2)$ & 9 \\
$e_3$ & $(1,1,2)$ & $(\omega,\omega^2,2)$ & $(\omega^2,\omega,2)$ & $(\omega,\omega^2,2\omega^2)$ & $(\omega^2,\omega,2\omega^2)$ & $(1,1,2\omega^2)$ & $(\omega^2,\omega,2\omega)$ & $(\omega,\omega^2,2\omega)$ & $(1,1,2\omega)$ & 9 \\
$b_{112}$ & $(2,-1,1)$ & $(2,-\omega,\omega^2)$ & $(2,-\omega^2,\omega)$ & $(2,-\omega,\omega)$ & $(2,-\omega^2,1)$ & $(2,-1,\omega^2)$ & $(2,-\omega^2,\omega^2)$ & $(2,-\omega,1)$ & $(2,-1,\omega)$ & 9 \\
$b_{212}$ & $(-1,2,1)$ & $(-1,2\omega,\omega^2)$ & $(-1,2\omega^2,\omega)$ & $(-1,2\omega,\omega)$ & $(-1,2\omega^2,1)$ & $(-1,2,\omega^2)$ & $(-1,2\omega^2,\omega^2)$ & $(-1,2\omega,1)$ & $(-1,2,\omega)$ & 9 \\
$b_{113}$ & $(2,1,-1)$ & $(2,\omega,-\omega^2)$ & $(2,\omega^2,-\omega)$ & $(2,\omega,-\omega)$ & $(2,\omega^2,-1)$ & $(2,1,-\omega^2)$ & $(2,\omega^2,-\omega^2)$ & $(2,\omega,-1)$ & $(2,1,-\omega)$ & 9 \\
$b_{313}$ & $(-1,1,2)$ & $(-1,\omega,2\omega^2)$ & $(-1,\omega^2,2\omega)$ & $(-1,\omega,2\omega)$ & $(-1,\omega^2,2)$ & $(-1,1,2\omega^2)$ & $(-1,\omega^2,2\omega^2)$ & $(-1,\omega,2)$ & $(-1,1,2\omega)$ & 9 \\
$b_{223}$ & $(1,2,-1)$ & $(1,2\omega,-\omega^2)$ & $(1,2\omega^2,-\omega)$ & $(1,2\omega,-\omega)$ & $(1,2\omega^2,-1)$ & $(1,2,-\omega^2)$ & $(1,2\omega^2,-\omega^2)$ & $(1,2\omega,-1)$ & $(1,2,-\omega)$ & 9 \\
$b_{323}$ & $(1,-1,2)$ & $(1,-\omega,2\omega^2)$ & $(1,-\omega^2,2\omega)$ & $(1,-\omega,2\omega)$ & $(1,-\omega^2,2)$ & $(1,-1,2\omega^2)$ & $(1,-\omega^2,2\omega^2)$ & $(1,-\omega,2)$ & $(1,-1,2\omega)$ & 9 \\
\end{tabular}
\end{ruledtabular}
\end{table*}

\section{Global inventory of 165 noncollinear rays and 130 orthogonal bases}
\label{sec:inventory}

A total of $165$ unique, noncollinear vectors were identified from the analysis of the
nine seed vectors \ensuremath{\sigma_1,\ldots,\sigma_9}
in Table~\ref{tab:vector_generation}.
These vectors, which form the basis for constructing orthogonal triples, are enumerated in Table~\ref{tab:noncollinear_vectors_6col}.
The coordinate system is based on the complex third root of unity, $\omega = \mathrm{e}^{2\pi \ii{}/3}$.

This vector set is not closed under the Hermitian cross product of Sec.~\ref{2025-yohscg} because, say,
$a_{11} \times d_{21} = (1,0,0) \times (1,-2,1) = -(0,1,2)$, which is not collinear with any of the $165$ rays.

\begin{table*}[htbp!]
\caption{
The complete set of $165$ projectively distinct rays, with six entries per row.  The rightmost index records the seed column in Table~\ref{tab:vector_generation}.  The $21$ rays common to all three 69-ray lineages are black.  The $48$ rays exclusive to $\mathcal{B}_1$, $\mathcal{B}_2$, and $\mathcal{B}_3$ are shown in red, green, and blue, respectively.  Each label is the rule symbol of Table~\ref{tab:vector_generation} followed by the seed index $j$; thus $b_{12}=b_1(\sigma_2)$, whereas the rule $b_{12}$ of Fig.~\ref{fig:yohs} appears as $b_{121},\ldots,b_{129}$.
}
\label{tab:noncollinear_vectors_6col}
\footnotesize 
\begin{ruledtabular}
\begin{tabular}{*{6}{c}} 
{$a_{11}=(1,0,0)$} & {$a_{21}=(0,1,0)$} & {$a_{31}=(0,0,1)$} & \textcolor{red}{$u_{1}=(1,1,1)$} & \textcolor{red}{$u_{2}=(1,\omega,\omega^2)$} & \textcolor{red}{$u_{3}=(1,\omega^2,\omega)$} \\
\textcolor{green!40!black}{$u_{4}=(1,\omega,\omega)$} & \textcolor{green!40!black}{$u_{5}=(1,\omega^2,1)$} & \textcolor{green!40!black}{$u_{6}=(1,1,\omega^2)$} & \textcolor{blue}{$u_{7}=(1,\omega^2,\omega^2)$} & \textcolor{blue}{$u_{8}=(1,\omega,1)$} & \textcolor{blue}{$u_{9}=(1,1,\omega)$} \\
{$b_{11}=(0,1,1)$} & {$b_{12}=(0,\omega,\omega^2)$} & {$b_{13}=(0,\omega^2,\omega)$} & {$b_{21}=(1,0,1)$} & {$b_{22}=(1,0,\omega^2)$} & {$b_{23}=(1,0,\omega)$} \\
{$b_{31}=(1,1,0)$} & {$b_{32}=(1,\omega,0)$} & {$b_{33}=(1,\omega^2,0)$} & {$c_{11}=(0,1,-1)$} & {$c_{12}=(0,\omega,-\omega^2)$} & {$c_{13}=(0,\omega^2,-\omega)$} \\
{$c_{21}=(-1,0,1)$} & {$c_{22}=(-\omega,0,1)$} & {$c_{23}=(-\omega^2,0,1)$} & {$c_{31}=(1,-1,0)$} & {$c_{32}=(\omega^2,-1,0)$} & {$c_{33}=(\omega,-1,0)$} \\
\textcolor{red}{$d_{11}=(-2,1,1)$} & \textcolor{red}{$d_{12}=(-2,\omega,\omega^2)$} & \textcolor{red}{$d_{13}=(-2,\omega^2,\omega)$} & \textcolor{green!40!black}{$d_{14}=(-2,\omega,\omega)$} & \textcolor{green!40!black}{$d_{15}=(-2,\omega^2,1)$} & \textcolor{green!40!black}{$d_{16}=(-2,1,\omega^2)$} \\
\textcolor{blue}{$d_{17}=(-2,\omega^2,\omega^2)$} & \textcolor{blue}{$d_{18}=(-2,\omega,1)$} & \textcolor{blue}{$d_{19}=(-2,1,\omega)$} & \textcolor{red}{$d_{21}=(1,-2,1)$} & \textcolor{red}{$d_{22}=(\omega^2,-2,\omega)$} & \textcolor{red}{$d_{23}=(\omega,-2,\omega^2)$} \\
\textcolor{green!40!black}{$d_{24}=(\omega^2,-2,1)$} & \textcolor{green!40!black}{$d_{25}=(\omega,-2,\omega)$} & \textcolor{green!40!black}{$d_{26}=(1,-2,\omega^2)$} & \textcolor{blue}{$d_{27}=(\omega,-2,1)$} & \textcolor{blue}{$d_{28}=(\omega^2,-2,\omega^2)$} & \textcolor{blue}{$d_{29}=(1,-2,\omega)$} \\
\textcolor{red}{$d_{31}=(1,1,-2)$} & \textcolor{red}{$d_{32}=(\omega,\omega^2,-2)$} & \textcolor{red}{$d_{33}=(\omega^2,\omega,-2)$} & \textcolor{green!40!black}{$d_{34}=(\omega^2,1,-2)$} & \textcolor{green!40!black}{$d_{35}=(1,\omega^2,-2)$} & \textcolor{green!40!black}{$d_{36}=(\omega,\omega,-2)$} \\
\textcolor{blue}{$d_{37}=(\omega,1,-2)$} & \textcolor{blue}{$d_{38}=(1,\omega,-2)$} & \textcolor{blue}{$d_{39}=(\omega^2,\omega^2,-2)$} & \textcolor{red}{$b_{121}=(1,1,-1)$} & \textcolor{red}{$b_{122}=(1,\omega,-\omega^2)$} & \textcolor{red}{$b_{123}=(1,\omega^2,-\omega)$} \\
\textcolor{green!40!black}{$b_{124}=(\omega,\omega^2,-\omega^2)$} & \textcolor{green!40!black}{$b_{125}=(\omega,1,-\omega)$} & \textcolor{green!40!black}{$b_{126}=(\omega,\omega,-1)$} & \textcolor{blue}{$b_{127}=(\omega^2,\omega,-\omega)$} & \textcolor{blue}{$b_{128}=(\omega^2,1,-\omega^2)$} & \textcolor{blue}{$b_{129}=(\omega^2,\omega^2,-1)$} \\
\textcolor{red}{$b_{131}=(1,-1,1)$} & \textcolor{red}{$b_{132}=(1,-\omega,\omega^2)$} & \textcolor{red}{$b_{133}=(1,-\omega^2,\omega)$} & \textcolor{green!40!black}{$b_{134}=(\omega,-\omega^2,\omega^2)$} & \textcolor{green!40!black}{$b_{135}=(\omega,-1,\omega)$} & \textcolor{green!40!black}{$b_{136}=(\omega,-\omega,1)$} \\
\textcolor{blue}{$b_{137}=(\omega^2,-\omega,\omega)$} & \textcolor{blue}{$b_{138}=(\omega^2,-1,\omega^2)$} & \textcolor{blue}{$b_{139}=(\omega^2,-\omega^2,1)$} & \textcolor{red}{$b_{231}=(-1,1,1)$} & \textcolor{red}{$b_{232}=(-1,\omega,\omega^2)$} & \textcolor{red}{$b_{233}=(-1,\omega^2,\omega)$} \\
\textcolor{green!40!black}{$b_{234}=(-\omega,\omega^2,\omega^2)$} & \textcolor{green!40!black}{$b_{235}=(-\omega,1,\omega)$} & \textcolor{green!40!black}{$b_{236}=(-\omega,\omega,1)$} & \textcolor{blue}{$b_{237}=(-\omega^2,\omega,\omega)$} & \textcolor{blue}{$b_{238}=(-\omega^2,1,\omega^2)$} & \textcolor{blue}{$b_{239}=(-\omega^2,\omega^2,1)$} \\
\textcolor{red}{$e_{11}=(2,1,1)$} & \textcolor{red}{$e_{12}=(2,\omega,\omega^2)$} & \textcolor{red}{$e_{13}=(2,\omega^2,\omega)$} & \textcolor{green!40!black}{$e_{14}=(2\omega^2,1,1)$} & \textcolor{green!40!black}{$e_{15}=(2\omega^2,\omega,\omega^2)$} & \textcolor{green!40!black}{$e_{16}=(2\omega^2,\omega^2,\omega)$} \\
\textcolor{blue}{$e_{17}=(2\omega,1,1)$} & \textcolor{blue}{$e_{18}=(2\omega,\omega^2,\omega)$} & \textcolor{blue}{$e_{19}=(2\omega,\omega,\omega^2)$} & \textcolor{red}{$e_{21}=(1,2,1)$} & \textcolor{red}{$e_{22}=(\omega^2,2,\omega)$} & \textcolor{red}{$e_{23}=(\omega,2,\omega^2)$} \\
\textcolor{green!40!black}{$e_{24}=(\omega,2\omega^2,\omega^2)$} & \textcolor{green!40!black}{$e_{25}=(1,2\omega^2,1)$} & \textcolor{green!40!black}{$e_{26}=(\omega^2,2\omega^2,\omega)$} & \textcolor{blue}{$e_{27}=(\omega^2,2\omega,\omega)$} & \textcolor{blue}{$e_{28}=(1,2\omega,1)$} & \textcolor{blue}{$e_{29}=(\omega,2\omega,\omega^2)$} \\
\textcolor{red}{$e_{31}=(1,1,2)$} & \textcolor{red}{$e_{32}=(\omega,\omega^2,2)$} & \textcolor{red}{$e_{33}=(\omega^2,\omega,2)$} & \textcolor{green!40!black}{$e_{34}=(\omega,\omega^2,2\omega^2)$} & \textcolor{green!40!black}{$e_{35}=(\omega^2,\omega,2\omega^2)$} & \textcolor{green!40!black}{$e_{36}=(1,1,2\omega^2)$} \\
\textcolor{blue}{$e_{37}=(\omega^2,\omega,2\omega)$} & \textcolor{blue}{$e_{38}=(\omega,\omega^2,2\omega)$} & \textcolor{blue}{$e_{39}=(1,1,2\omega)$} & \textcolor{red}{$b_{1121}=(2,-1,1)$} & \textcolor{red}{$b_{1122}=(2,-\omega,\omega^2)$} & \textcolor{red}{$b_{1123}=(2,-\omega^2,\omega)$} \\
\textcolor{green!40!black}{$b_{1124}=(2,-\omega,\omega)$} & \textcolor{green!40!black}{$b_{1125}=(2,-\omega^2,1)$} & \textcolor{green!40!black}{$b_{1126}=(2,-1,\omega^2)$} & \textcolor{blue}{$b_{1127}=(2,-\omega^2,\omega^2)$} & \textcolor{blue}{$b_{1128}=(2,-\omega,1)$} & \textcolor{blue}{$b_{1129}=(2,-1,\omega)$} \\
\textcolor{red}{$b_{2121}=(-1,2,1)$} & \textcolor{red}{$b_{2122}=(-1,2\omega,\omega^2)$} & \textcolor{red}{$b_{2123}=(-1,2\omega^2,\omega)$} & \textcolor{green!40!black}{$b_{2124}=(-1,2\omega,\omega)$} & \textcolor{green!40!black}{$b_{2125}=(-1,2\omega^2,1)$} & \textcolor{green!40!black}{$b_{2126}=(-1,2,\omega^2)$} \\
\textcolor{blue}{$b_{2127}=(-1,2\omega^2,\omega^2)$} & \textcolor{blue}{$b_{2128}=(-1,2\omega,1)$} & \textcolor{blue}{$b_{2129}=(-1,2,\omega)$} & \textcolor{red}{$b_{1131}=(2,1,-1)$} & \textcolor{red}{$b_{1132}=(2,\omega,-\omega^2)$} & \textcolor{red}{$b_{1133}=(2,\omega^2,-\omega)$} \\
\textcolor{green!40!black}{$b_{1134}=(2,\omega,-\omega)$} & \textcolor{green!40!black}{$b_{1135}=(2,\omega^2,-1)$} & \textcolor{green!40!black}{$b_{1136}=(2,1,-\omega^2)$} & \textcolor{blue}{$b_{1137}=(2,\omega^2,-\omega^2)$} & \textcolor{blue}{$b_{1138}=(2,\omega,-1)$} & \textcolor{blue}{$b_{1139}=(2,1,-\omega)$} \\
\textcolor{red}{$b_{3131}=(-1,1,2)$} & \textcolor{red}{$b_{3132}=(-1,\omega,2\omega^2)$} & \textcolor{red}{$b_{3133}=(-1,\omega^2,2\omega)$} & \textcolor{green!40!black}{$b_{3134}=(-1,\omega,2\omega)$} & \textcolor{green!40!black}{$b_{3135}=(-1,\omega^2,2)$} & \textcolor{green!40!black}{$b_{3136}=(-1,1,2\omega^2)$} \\
\textcolor{blue}{$b_{3137}=(-1,\omega^2,2\omega^2)$} & \textcolor{blue}{$b_{3138}=(-1,\omega,2)$} & \textcolor{blue}{$b_{3139}=(-1,1,2\omega)$} & \textcolor{red}{$b_{2231}=(1,2,-1)$} & \textcolor{red}{$b_{2232}=(1,2\omega,-\omega^2)$} & \textcolor{red}{$b_{2233}=(1,2\omega^2,-\omega)$} \\
\textcolor{green!40!black}{$b_{2234}=(1,2\omega,-\omega)$} & \textcolor{green!40!black}{$b_{2235}=(1,2\omega^2,-1)$} & \textcolor{green!40!black}{$b_{2236}=(1,2,-\omega^2)$} & \textcolor{blue}{$b_{2237}=(1,2\omega^2,-\omega^2)$} & \textcolor{blue}{$b_{2238}=(1,2\omega,-1)$} & \textcolor{blue}{$b_{2239}=(1,2,-\omega)$} \\
\textcolor{red}{$b_{3231}=(1,-1,2)$} & \textcolor{red}{$b_{3232}=(1,-\omega,2\omega^2)$} & \textcolor{red}{$b_{3233}=(1,-\omega^2,2\omega)$} & \textcolor{green!40!black}{$b_{3234}=(1,-\omega,2\omega)$} & \textcolor{green!40!black}{$b_{3235}=(1,-\omega^2,2)$} & \textcolor{green!40!black}{$b_{3236}=(1,-1,2\omega^2)$} \\
\textcolor{blue}{$b_{3237}=(1,-\omega^2,2\omega^2)$} & \textcolor{blue}{$b_{3238}=(1,-\omega,2)$} & \textcolor{blue}{$b_{3239}=(1,-1,2\omega)$} & & & \\
\end{tabular}
\end{ruledtabular}
\end{table*}

\begin{table*}[htbp!]
\caption{
The complete list of $130$ orthogonal bases supported by the $165$ rays.
The invariant provenance counts are stated in Proposition~\ref{prop:provenance}.
}
\label{tab:orthogonal_bases}
\begin{ruledtabular}
\begin{tabular}{*{6}{c}}

{\{$a_{11}$, $a_{21}$, $a_{31}$\}} & {\{$a_{11}$, $b_{11}$, $c_{11}$\}} & {\{$a_{11}$, $b_{12}$, $c_{12}$\}} & {\{$a_{11}$, $b_{13}$, $c_{13}$\}} & {\{$a_{21}$, $b_{21}$, $c_{21}$\}} & {\{$a_{21}$, $b_{22}$, $c_{22}$\}} \\
{\{$a_{21}$, $b_{23}$, $c_{23}$\}} & {\{$a_{31}$, $b_{31}$, $c_{31}$\}} & {\{$a_{31}$, $b_{32}$, $c_{32}$\}} & {\{$a_{31}$, $b_{33}$, $c_{33}$\}} & \{$u_{1}$, $u_{2}$, $u_{3}$\} & \{$u_{1}$, $c_{11}$, $d_{11}$\} \\
\{$u_{1}$, $c_{21}$, $d_{21}$\} & \{$u_{1}$, $c_{31}$, $d_{31}$\} & \{$u_{2}$, $c_{12}$, $d_{12}$\} & \{$u_{2}$, $c_{22}$, $d_{22}$\} & \{$u_{2}$, $c_{32}$, $d_{32}$\} & \{$u_{3}$, $c_{13}$, $d_{13}$\} \\
\{$u_{3}$, $c_{23}$, $d_{23}$\} & \{$u_{3}$, $c_{33}$, $d_{33}$\} & \{$u_{4}$, $u_{5}$, $u_{6}$\} & \{$u_{4}$, $c_{11}$, $d_{14}$\} & \{$u_{4}$, $c_{23}$, $d_{24}$\} & \{$u_{4}$, $c_{32}$, $d_{34}$\} \\
\{$u_{5}$, $c_{12}$, $d_{15}$\} & \{$u_{5}$, $c_{21}$, $d_{25}$\} & \{$u_{5}$, $c_{33}$, $d_{35}$\} & \{$u_{6}$, $c_{13}$, $d_{16}$\} & \{$u_{6}$, $c_{22}$, $d_{26}$\} & \{$u_{6}$, $c_{31}$, $d_{36}$\} \\
\{$u_{7}$, $u_{8}$, $u_{9}$\} & \{$u_{7}$, $c_{11}$, $d_{17}$\} & \{$u_{7}$, $c_{22}$, $d_{27}$\} & \{$u_{7}$, $c_{33}$, $d_{37}$\} & \{$u_{8}$, $c_{13}$, $d_{18}$\} & \{$u_{8}$, $c_{21}$, $d_{28}$\} \\
\{$u_{8}$, $c_{32}$, $d_{38}$\} & \{$u_{9}$, $c_{12}$, $d_{19}$\} & \{$u_{9}$, $c_{23}$, $d_{29}$\} & \{$u_{9}$, $c_{31}$, $d_{39}$\} & \{$b_{11}$, $b_{121}$, $b_{1121}$\} & \{$b_{11}$, $b_{124}$, $b_{1124}$\} \\
\{$b_{11}$, $b_{127}$, $b_{1127}$\} & \{$b_{11}$, $b_{131}$, $b_{1131}$\} & \{$b_{11}$, $b_{134}$, $b_{1134}$\} & \{$b_{11}$, $b_{137}$, $b_{1137}$\} & \{$b_{12}$, $b_{122}$, $b_{1122}$\} & \{$b_{12}$, $b_{125}$, $b_{1125}$\} \\
\{$b_{12}$, $b_{129}$, $b_{1129}$\} & \{$b_{12}$, $b_{132}$, $b_{1132}$\} & \{$b_{12}$, $b_{135}$, $b_{1135}$\} & \{$b_{12}$, $b_{139}$, $b_{1139}$\} & \{$b_{13}$, $b_{123}$, $b_{1123}$\} & \{$b_{13}$, $b_{126}$, $b_{1126}$\} \\
\{$b_{13}$, $b_{128}$, $b_{1128}$\} & \{$b_{13}$, $b_{133}$, $b_{1133}$\} & \{$b_{13}$, $b_{136}$, $b_{1136}$\} & \{$b_{13}$, $b_{138}$, $b_{1138}$\} & \{$b_{21}$, $b_{121}$, $b_{2121}$\} & \{$b_{21}$, $b_{125}$, $b_{2125}$\} \\
\{$b_{21}$, $b_{128}$, $b_{2128}$\} & \{$b_{21}$, $b_{231}$, $b_{2231}$\} & \{$b_{21}$, $b_{235}$, $b_{2235}$\} & \{$b_{21}$, $b_{238}$, $b_{2238}$\} & \{$b_{22}$, $b_{122}$, $b_{2122}$\} & \{$b_{22}$, $b_{126}$, $b_{2126}$\} \\
\{$b_{22}$, $b_{127}$, $b_{2127}$\} & \{$b_{22}$, $b_{232}$, $b_{2232}$\} & \{$b_{22}$, $b_{236}$, $b_{2236}$\} & \{$b_{22}$, $b_{237}$, $b_{2237}$\} & \{$b_{23}$, $b_{123}$, $b_{2123}$\} & \{$b_{23}$, $b_{124}$, $b_{2124}$\} \\
\{$b_{23}$, $b_{129}$, $b_{2129}$\} & \{$b_{23}$, $b_{233}$, $b_{2233}$\} & \{$b_{23}$, $b_{234}$, $b_{2234}$\} & \{$b_{23}$, $b_{239}$, $b_{2239}$\} & \{$b_{31}$, $b_{131}$, $b_{3131}$\} & \{$b_{31}$, $b_{136}$, $b_{3136}$\} \\
\{$b_{31}$, $b_{139}$, $b_{3139}$\} & \{$b_{31}$, $b_{231}$, $b_{3231}$\} & \{$b_{31}$, $b_{236}$, $b_{3236}$\} & \{$b_{31}$, $b_{239}$, $b_{3239}$\} & \{$b_{32}$, $b_{132}$, $b_{3132}$\} & \{$b_{32}$, $b_{134}$, $b_{3134}$\} \\
\{$b_{32}$, $b_{138}$, $b_{3138}$\} & \{$b_{32}$, $b_{232}$, $b_{3232}$\} & \{$b_{32}$, $b_{234}$, $b_{3234}$\} & \{$b_{32}$, $b_{238}$, $b_{3238}$\} & \{$b_{33}$, $b_{133}$, $b_{3133}$\} & \{$b_{33}$, $b_{135}$, $b_{3135}$\} \\
\{$b_{33}$, $b_{137}$, $b_{3137}$\} & \{$b_{33}$, $b_{233}$, $b_{3233}$\} & \{$b_{33}$, $b_{235}$, $b_{3235}$\} & \{$b_{33}$, $b_{237}$, $b_{3237}$\} & \{$c_{11}$, $b_{231}$, $e_{11}$\} & \{$c_{11}$, $b_{234}$, $e_{14}$\} \\
\{$c_{11}$, $b_{237}$, $e_{17}$\} & \{$c_{12}$, $b_{232}$, $e_{12}$\} & \{$c_{12}$, $b_{235}$, $e_{15}$\} & \{$c_{12}$, $b_{239}$, $e_{19}$\} & \{$c_{13}$, $b_{233}$, $e_{13}$\} & \{$c_{13}$, $b_{236}$, $e_{16}$\} \\
\{$c_{13}$, $b_{238}$, $e_{18}$\} & \{$c_{21}$, $b_{131}$, $e_{21}$\} & \{$c_{21}$, $b_{135}$, $e_{25}$\} & \{$c_{21}$, $b_{138}$, $e_{28}$\} & \{$c_{22}$, $b_{132}$, $e_{22}$\} & \{$c_{22}$, $b_{136}$, $e_{26}$\} \\
\{$c_{22}$, $b_{137}$, $e_{27}$\} & \{$c_{23}$, $b_{133}$, $e_{23}$\} & \{$c_{23}$, $b_{134}$, $e_{24}$\} & \{$c_{23}$, $b_{139}$, $e_{29}$\} & \{$c_{31}$, $b_{121}$, $e_{31}$\} & \{$c_{31}$, $b_{126}$, $e_{36}$\} \\
\{$c_{31}$, $b_{129}$, $e_{39}$\} & \{$c_{32}$, $b_{122}$, $e_{32}$\} & \{$c_{32}$, $b_{124}$, $e_{34}$\} & \{$c_{32}$, $b_{128}$, $e_{38}$\} & \{$c_{33}$, $b_{123}$, $e_{33}$\} & \{$c_{33}$, $b_{125}$, $e_{35}$\} \\
\{$c_{33}$, $b_{127}$, $e_{37}$\} & \{$b_{121}$, $b_{122}$, $b_{123}$\} & \{$b_{124}$, $b_{125}$, $b_{126}$\} & \{$b_{127}$, $b_{128}$, $b_{129}$\} & \{$b_{131}$, $b_{132}$, $b_{133}$\} & \{$b_{134}$, $b_{135}$, $b_{136}$\} \\
\{$b_{137}$, $b_{138}$, $b_{139}$\} & \{$b_{231}$, $b_{232}$, $b_{233}$\} & \{$b_{234}$, $b_{235}$, $b_{236}$\} & \{$b_{237}$, $b_{238}$, $b_{239}$\} & & \\
\end{tabular}
\end{ruledtabular}
\end{table*}

\paragraph*{Symmetric provenance decomposition.}

For $r=1,2,3$, let $\mathcal{S}_r$ be the projective ray set generated from the three seeds in $\mathcal{B}_r$, and let $\mathcal{E}_r$ be its set of orthogonal triples.

\begin{proposition}[Ray and context inventory]
\label{prop:provenance}
The generated families satisfy
\[
|\mathcal{S}_r|=69,\qquad
|\mathcal{S}_1\cap\mathcal{S}_2\cap\mathcal{S}_3|=21,
\qquad |\mathcal{S}_r\setminus(\mathcal{S}_s\cup\mathcal{S}_t)|=48,
\]
where $\{r,s,t\}=\{1,2,3\}$, and no ray belongs to exactly two lineages.  Their union has $165$
rays.  Likewise, each $\mathcal{E}_r$ contains $50$ contexts, $10$ contexts are common to all three lineages, and each lineage contributes $40$ further contexts.  The union therefore contains $10+3\cdot40=130$ contexts, and these are all the orthogonal triples among the $165$ rays; in particular, no context combines exclusive rays of two different lineages.  Exactly four contexts in each lineage consist entirely of lineage-exclusive rays.
\end{proposition}

\begin{proof}
Each nonzero generated vector is divided by its first nonzero coordinate to obtain a canonical projective representative.  Equality is then decided using $1+\omega+\omega^2=0$.  This gives $69$ representatives per lineage and the overlap counts above.  Pairwise Hermitian inner products are evaluated exactly in $\mathbb{Q}(\omega)$; triangles in the resulting orthogonality graph are precisely the contexts.  The exhaustive search over all $165$ rays gives $130$ triangles, which coincide with the union of the three lineage sets of $50$ contexts each.  Finally, direct enumeration of maps $s:V\to\{0,1\}$ with one $1$ per context gives no two-valued state for any of the three $69$-ray, $50$-context lineages.
\end{proof}

The symmetric provenance statement depends on the complete sets of generating lineages and is independent of the display labels assigned to the rays.


\section{Higher-dimensional forcing gadgets}
\label{2025-hdfg}

In $\mathcal{P}(\mathbb{C}^3)$, two independent rays determine a unique orthogonal ray.  For $D\geq4$, their orthogonal complement has dimension $D-2$, so additional connector contexts are required to obtain a rigid finite construction.

\begin{definition}[Unbiased vector]
\label{def:unbiased}
A normalized vector $u$ is maximally unbiased relative to an orthonormal basis $\mathcal{B}=\{e_1,\ldots,e_D\}$ if
\[
|\langle e_j,u\rangle|^2=1/D\quad (j=1,\ldots,D).
\]
For $u=(x_1,\ldots,x_D)$ in the computational basis, this is equivalent to \ensuremath{|x_1|=\cdots=|x_D|}.
\end{definition}

The $D=4$ and $D=5$ constructions have three logically separate parts: (i) a scaffold of coordinate-supported minors, (ii) connector rays whose orthogonality to the center gives equations in $|x_j|^2$, and (iii) completion vectors that turn the required orthogonal pairs into full contexts.  The propositions below state the forcing constraints; coordinate inventories and exact enumeration checks are given afterward.

\subsection{Notation and conventions}
Fix an unknown center vector $u = (x_1,\dots,x_D) \in \mathbb{C}^D$ with $x_j \ne 0$ for all $j$, and use the standard Hermitian inner product
\[
\langle v,w\rangle = \sum_m \overline{v_m}\, w_m.
\]
We use vectors to represent the one-dimensional subspaces they span and freely scale them when forming blocks,
with the understanding that they must be normalized to form a literal orthonormal basis.
From here on, $e_1,\ldots,e_D$ denote the computational basis of $\mathbb{C}^D$; they are unrelated to the rays $e_i$ of Secs.~\ref{2025-yohscg} and~\ref{sec:inventory}.

\subsection{An explicit gadget for \texorpdfstring{$D=4$}{D=4}}

First, take the computational (Cartesian) basis
\begin{align*}
e_1&=(1,0,0,0), &
e_2 =(0,1,0,0),\\
e_3&=(0,0,1,0), &
e_4 =(0,0,0,1),
\end{align*}
and place the context $\mathcal{B}_{\mathrm{comp}}=\{e_1,e_2,e_3,e_4\}$ on the rim of Fig.~\ref{fig:peres}(a).

In the second construction stage define, for each ordered pair $(i,j)$ with $1\le i<j\le 4$, pair minors from the Levi-Civita contraction $(v_{ij})_m=\sum_{\ell=1}^4\varepsilon_{mij\ell}\,\overline{x_\ell}$:
\begin{align*}
v_{12} &= (0,0,\overline{x_4},-\overline{x_3}), &
v_{13}  = (0,-\overline{x_4},0,\overline{x_2}), \\
v_{14} &= (0,\overline{x_3},-\overline{x_2},0), &
v_{23}  = (\overline{x_4},0,0,-\overline{x_1}), \\
v_{24} &= (-\overline{x_3},0,\overline{x_1},0), &
v_{34}  = (\overline{x_2},-\overline{x_1},0,0).
\end{align*}
Each $v_{ij}$ vanishes in coordinates $i,j$; the complex conjugation cancels the conjugation in the Hermitian inner product, giving $\langle v_{ij},u\rangle=0$ by direct cancellation for all $u\in\mathbb{C}^4$.

In the third step, choose a vector $w_{ij}$ with the same support as $v_{ij}$ and orthogonal to it, that is, $\langle v_{ij},w_{ij}\rangle=0$:
\begin{align*}
w_{12}&=(0,0,x_3,x_4), &
w_{13} =(0,x_2,0,x_4), \\
w_{14}&=(0,x_2,x_3,0), &
w_{23} =(x_1,0,0,x_4), \\
w_{24}&=(x_1,0,x_3,0), &
w_{34} =(x_1,x_2,0,0).
\end{align*}
For each of the $\binom{4}{2}=6$ pairs $\{i,j\}$, the set
\[
\mathcal{B}_{ij}=\{e_i, e_j, v_{ij}, w_{ij}\}
\]
is an orthogonal basis (block, context).

In a fourth step, define three connector vectors
\begin{align*}
v_{1234}&=(x_1,x_2,-x_3,-x_4),\\
v_{1324}&=(x_1,-x_2,x_3,-x_4),\\
v_{1423}&=(x_1,-x_2,-x_3,x_4),
\end{align*}
and require the three connector contexts of Fig.~\ref{fig:peres}(a),
\begin{align*}
\mathcal{C}_{(a)}&=\{u,v_{1234},v_{34},v_{12}\},\\
\mathcal{C}_{(b)}&=\{u,v_{1324},v_{24},v_{13}\},\\
\mathcal{C}_{(c)}&=\{u,v_{1423},v_{23},v_{14}\}.
\end{align*}
For every $u\in(\mathbb{C}\setminus\{0\})^4$, the two pair minors in each quadruple have disjoint support and are already orthogonal to both $u$ and the corresponding connector vector (for instance, $\langle v_{34},v_{1234}\rangle=x_2x_1-x_1x_2=0$ and $\langle v_{12},v_{1234}\rangle=-x_4x_3+x_3x_4=0$).  Hence $\mathcal{C}_{(a)}$, $\mathcal{C}_{(b)}$, and $\mathcal{C}_{(c)}$ are orthogonal bases precisely when $u$ is orthogonal to $v_{1234}$, $v_{1324}$, and $v_{1423}$, respectively.

\begin{proposition}[$D=4$ forcing constraint]
\label{prop:D4forcing}
With the connector coordinates given above, membership of $u$ and the three connector rays in these contexts forces a normalized center vector $u$ to be maximally unbiased relative to $\mathcal{B}_{\mathrm{comp}}$.
\end{proposition}

\begin{proof}
The three required inner products vanish exactly when
\begin{align}
|x_1|^2+|x_2|^2&=|x_3|^2+|x_4|^2, \label{eq:D4a}\\
|x_1|^2+|x_3|^2&=|x_2|^2+|x_4|^2, \label{eq:D4b}\\
|x_1|^2+|x_4|^2&=|x_2|^2+|x_3|^2. \label{eq:D4c}
\end{align}
Subtracting pairs of equations gives $|x_2|^2=|x_3|^2=|x_4|^2$; substitution gives $|x_1|^2=|x_2|^2$.  Since the moduli are nonnegative, \ensuremath{|x_1|=|x_2|=|x_3|=|x_4|}.  After normalization, every squared modulus is $1/4$.
\end{proof}

\subsubsection{Automatic and equality-induced contexts}
Orthogonalities due only to disjoint support, such as $v_{12}\perp v_{34}$, do not constrain $u$.  For the concrete 20-ray representation with $u=(1,1,1,1)$, the computational context, six pair contexts, and three connector contexts give 10 imposed contexts.  An exhaustive orthogonality search finds exactly seven further contexts.  Three follow from disjoint support:
\begin{align*}
\mathcal{B}_{1234}&=\{v_{12},w_{12},v_{34},w_{34}\},\\
\mathcal{B}_{1324}&=\{v_{13},w_{13},v_{24},w_{24}\},\\
\mathcal{B}_{1423}&=\{v_{14},w_{14},v_{23},w_{23}\}.
\end{align*}
The remaining four use the equal-modulus relations from Proposition~\ref{prop:D4forcing}:
\begin{align*}
\mathcal{C}_0&=\{u,v_{1234},v_{1324},v_{1423}\},\\
\mathcal{E}_1&=\{w_{12},w_{34},v_{1324},v_{1423}\},\\
\mathcal{E}_2&=\{w_{13},w_{24},v_{1234},v_{1423}\},\\
\mathcal{E}_3&=\{w_{14},w_{23},v_{1234},v_{1324}\}.
\end{align*}
Hence the complete orthogonality hypergraph of the concrete $20$-ray set has $17$ contexts, not $13$.  Exact enumeration gives $16$ two-valued states, which are separating and unital; the complete $20$-ray hypergraph is therefore not a KS configuration.

\begin{figure*}[ht]
\centering
\begin{tabular}{ccc}
\begin{tikzpicture}[thick, scale=0.35]
    \colorlet{colorCircle}{gray}
    \colorlet{colorE1E2}{red} \colorlet{colorE2E3}{orange}
    \colorlet{colorE3E4}{blue} \colorlet{colorE4E1}{cyan}
    \colorlet{colorV12V34}{green} \colorlet{colorV23V41}{lime!70!black}
    \colorlet{colorCurve13}{magenta} \colorlet{colorCurve24}{violet}
    \colorlet{colorV13V24}{purple}
    \colorlet{colorHiddenA}{brown} \colorlet{colorHiddenB}{teal}
    \colorlet{colorEllipse}{black}   
    \colorlet{darkgreen}{green!40!black}

    \def\R{10} \def\bendangle{30}

    \foreach \angle/\i in {45/1, 315/2, 225/3, 135/4} { \coordinate (e\i) at (\angle:\R); }
    \foreach \i/\j in {1/2, 2/3, 3/4, 4/1} {
        \coordinate (v\i\j) at ($(e\i)!1/3!(e\j)$);
        \coordinate (w\i\j) at ($(e\i)!2/3!(e\j)$);
    }
    \coordinate (v13) at ($(e1)!1/4!(e3)$);
    \coordinate (v24) at ($(e2)!1/4!(e4)$);
    \coordinate (w13) at ($(e1)!3/4!(e3)$);
    \coordinate (w24) at ($(e2)!3/4!(e4)$);

    \path[name path=curve-v12-v34] (v34) -- (v12);
    \path[name path=curve-v13-v24] (v13) -- (v24);
    \path[name intersections={of=curve-v12-v34 and curve-v13-v24, by=u}];

    \coordinate (g12) at ($(u)!0.5!(v12)$);
    \coordinate (g13) at ($(u)!0.5!(v13)$);
    \coordinate (g23) at (325:\R*.6); 
    \path (e1) (v13) (w13) (e3);
    \path (e2) (v24) (w24) (e4);

    \draw[colorCircle] (0,0) circle (\R);
    \draw[colorE1E2] (e1) -- (e2); \draw[colorE2E3] (e2) -- (e3);
    \draw[colorE3E4] (e3) -- (e4); \draw[colorE4E1] (e4) -- (e1);
    \draw[colorCurve13] (e1) -- (e3);
    \draw[colorCurve24] (e2) -- (e4);

    \draw[colorHiddenA] (v41) -- (w23); \draw[colorHiddenA] (w41) -- (v23);
    \draw[colorHiddenB] (v12) -- (w34); \draw[colorHiddenB] (w12) -- (v34);
    \draw[colorEllipse] (0,0) circle (\R/2);
    \draw[colorHiddenA] let \p1=($(w41)-(v41)$), \n1={atan2(\y1,\x1)}, \n2={veclen(\x1,\y1)/2} in (w41) arc (\n1:\n1+180:\n2);
    \draw[colorHiddenA] let \p1=($(w23)-(v23)$), \n1={atan2(\y1,\x1)}, \n2={veclen(\x1,\y1)/2} in (w23) arc (\n1:\n1+180:\n2);
    \draw[colorHiddenB] let \p1=($(w12)-(v12)$), \n1={atan2(\y1,\x1)}, \n2={veclen(\x1,\y1)/2} in (w12) arc (\n1:\n1+180:\n2);
    \draw[colorHiddenB] let \p1=($(w34)-(v34)$), \n1={atan2(\y1,\x1)}, \n2={veclen(\x1,\y1)/2} in (w34) arc (\n1:\n1+180:\n2);

    \draw[colorV12V34] (v34) -- (v12);

    \draw[colorV23V41] plot[smooth, tension=0.5] coordinates {(v41) (u) (g23) (v23)};

    \draw[darkgreen, dotted, line width=1.5pt] plot[smooth, tension=0.5] coordinates {(w41) (g12) (w23)};
    \path[name path=curve-w41-w23] plot[smooth, tension=0.5] coordinates {(w41) (g12) (w23)};
    \path[name intersections={of=curve-w41-w23 and curve-v13-v24, by=v1324}];
    \draw[colorV13V24] (v24) -- (v13);
    \node[circle, fill=darkgreen, inner sep=2.5pt] at (v1324) {};
    \node[circle, fill=colorV13V24, inner sep=1.5pt, label={left:$v_{1324}$}] at (v1324) {};

    \node[circle, fill=darkgreen, inner sep=2.5pt] at (g12) {};
    \node[circle, fill=colorV12V34, inner sep=1.5pt, label={below right:$v_{1234}$}] at (g12) {};
    \node[circle, fill=colorV23V41, inner sep=2.5pt, label={below right:$v_{1423}$}] at (g23) {};
    \node[circle, fill=colorV12V34, inner sep=3pt, label={left:$u$}] at (u) {}; \node[circle, fill=colorV23V41, inner sep=2pt] at (u) {}; \node[circle, fill=colorV13V24, inner sep=1pt] at (u) {};
    \node[circle, fill=colorCurve13, inner sep=3pt, label={right:$v_{13}$}] at (v13) {}; \node[circle, fill=colorV13V24, inner sep=2pt] at (v13) {}; \node[circle, fill=colorEllipse, inner sep=1pt] at (v13) {};
    \node[circle, fill=colorCurve24, inner sep=3pt, label={left:$v_{24}$}] at (v24) {}; \node[circle, fill=colorV13V24, inner sep=2pt] at (v24) {}; \node[circle, fill=colorEllipse, inner sep=1pt] at (v24) {};
    \node[circle, fill=colorCurve13, inner sep=2.5pt, label={left:$w_{13}$}] at (w13) {}; \node[circle, fill=colorEllipse, inner sep=1.25pt] at (w13) {};
    \node[circle, fill=colorCurve24, inner sep=2.5pt, label={left:$w_{24}$}] at (w24) {}; \node[circle, fill=colorEllipse, inner sep=1.25pt] at (w24) {};
    \node[circle, fill=darkgreen, inner sep=3.5pt] at (w41) {};
    \node[circle, fill=colorE4E1, inner sep=2.5pt, label={above right:$w_{14}$}] at (w41) {}; \node[circle, fill=colorHiddenA, inner sep=1.25pt] at (w41) {};
    \node[circle, fill=darkgreen, inner sep=3.5pt] at (w23) {};
    \node[circle, fill=colorE2E3, inner sep=2.5pt, label={below left:$w_{23}$}] at (w23) {}; \node[circle, fill=colorHiddenA, inner sep=1.25pt] at (w23) {};
    \node[circle, fill=colorE4E1, inner sep=3pt, label={above left:$v_{14}$}] at (v41) {}; \node[circle, fill=colorV23V41, inner sep=2pt] at (v41) {}; \node[circle, fill=colorHiddenA, inner sep=1pt] at (v41) {};
    \node[circle, fill=colorE2E3, inner sep=3pt, label={below right:$v_{23}$}] at (v23) {}; \node[circle, fill=colorV23V41, inner sep=2pt] at (v23) {}; \node[circle, fill=colorHiddenA, inner sep=1pt] at (v23) {};
    \node[circle, fill=colorE1E2, inner sep=3pt, label={above right:$v_{12}$}] at (v12) {}; \node[circle, fill=colorV12V34, inner sep=2pt] at (v12) {}; \node[circle, fill=colorHiddenB, inner sep=1pt] at (v12) {};
    \node[circle, fill=colorE1E2, inner sep=2.5pt, label={below right:$w_{12}$}] at (w12) {}; \node[circle, fill=colorHiddenB, inner sep=1.25pt] at (w12) {};
    \node[circle, fill=colorE3E4, inner sep=3pt, label={below left:$v_{34}$}] at (v34) {}; \node[circle, fill=colorV12V34, inner sep=2pt] at (v34) {}; \node[circle, fill=colorHiddenB, inner sep=1pt] at (v34) {};
    \node[circle, fill=colorE3E4, inner sep=2.5pt, label={above left:$w_{34}$}] at (w34) {}; \node[circle, fill=colorHiddenB, inner sep=1.25pt] at (w34) {};
    \node[circle, fill=colorCircle, inner sep=4pt, label={45:$e_1$}] at (e1) {}; \node[circle, fill=colorE4E1, inner sep=3pt] at (e1) {}; \node[circle, fill=colorE1E2, inner sep=2pt] at (e1) {}; \node[circle, fill=colorCurve13, inner sep=1pt] at (e1) {};
    \node[circle, fill=colorCircle, inner sep=4pt, label={315:$e_2$}] at (e2) {}; \node[circle, fill=colorE1E2, inner sep=3pt] at (e2) {}; \node[circle, fill=colorE2E3, inner sep=2pt] at (e2) {}; \node[circle, fill=colorCurve24, inner sep=1pt] at (e2) {};
    \node[circle, fill=colorCircle, inner sep=4pt, label={225:$e_3$}] at (e3) {}; \node[circle, fill=colorE2E3, inner sep=3pt] at (e3) {}; \node[circle, fill=colorE3E4, inner sep=2pt] at (e3) {}; \node[circle, fill=colorCurve13, inner sep=1pt] at (e3) {};
    \node[circle, fill=colorCircle, inner sep=4pt, label={135:$e_4$}] at (e4) {}; \node[circle, fill=colorE3E4, inner sep=3pt] at (e4) {}; \node[circle, fill=colorE4E1, inner sep=2pt] at (e4) {}; \node[circle, fill=colorCurve24, inner sep=1pt] at (e4) {};

\end{tikzpicture}
&\qquad \qquad \qquad &
\begin{tikzpicture}  [ultra thick, scale=0.6]

    \colorlet{colorE1E2}{red} \colorlet{colorE4E1}{cyan}
    \colorlet{colorV12V34}{green} \colorlet{colorV13V24}{purple}
    \colorlet{colorCurve24}{violet} \colorlet{lightgray}{gray!50!}
    \colorlet{darkgreen}{green!40!black}

    \tikzstyle{every path}=[line width=1pt]
    \tikzstyle{c2}=[circle,inner sep=3pt,minimum size=9pt]
    \tikzstyle{c1}=[circle,inner sep=0pt,minimum size=5pt]

    \path
              (240:5) coordinate(1) (-0.833,-4.33) coordinate(2)
              (0.833,-4.33) coordinate(3) (300:5) coordinate(4)
              (3.33,-2.88) coordinate(5) (4.167,-1.44) coordinate(6)
              (0:5) coordinate(7) (4.167,1.44) coordinate(8)
              (3.33,2.88) coordinate(9) (60:5) coordinate(10)
              (0.833,4.33) coordinate(11) (-0.833,4.33) coordinate(12)
              (120:5) coordinate(13) (-3.33,2.88) coordinate(14)
              (-4.167,1.44) coordinate(15) (180:5) coordinate(16)
              (-4.167,-1.44) coordinate(17) (-3.33,-2.88) coordinate(18);

    \draw [color=lightgray] (7) -- (8) -- (9)-- (10);
    \draw [color=lightgray] (10) -- (11) -- (12) -- (13);
    \draw [color=lightgray] (9) -- (2); \draw [color=lightgray] (11) -- (18);
    \draw [rotate=240,color=lightgray] (9) arc (90:270:2 and 1.44); \draw[rotate=60,color=lightgray] (18) arc (90:270:2 and 1.44);
    \draw [color=darkgreen] (12) -- (5); \draw [color=darkgreen] (14) -- (3);
    \draw[rotate=300,color=darkgreen] (12) arc (90:270:2 and 1.44); \draw[rotate=120,color=darkgreen] (3) arc (90:270:2 and 1.44);
    \draw [color=colorE4E1] (13) -- (14) -- (15) -- (16);
    \draw [color=colorE1E2] (16) -- (17) -- (18) -- (1);
    \draw [color=colorV12V34] (1) -- (2) -- (3) -- (4);
    \draw [color=colorV13V24] (4) -- (5) -- (6) -- (7);
    \draw [color=colorCurve24] (8) -- (15); \draw [color=colorCurve24](17) -- (6);
    \draw [color=colorCurve24] (8) arc (450:270:2 and 1.44); \draw [color=colorCurve24] (15) arc (90:270:2 and 1.44);

    \draw (1) coordinate[c2,fill=colorV12V34]; \draw (1) coordinate[c1,fill=colorE1E2,label=85:$v_{12}$];
    \draw (2) coordinate[c2,fill=colorV12V34]; \draw (2) coordinate[c1,fill=lightgray,label={[label distance=2mm,align=center]270:$v_{34}$}];
    \draw (3) coordinate[c2,fill=colorV12V34]; \draw (3) coordinate[c1,fill=darkgreen,label={[label distance=2mm,align=center]270:$v_{1234}$}];
    \draw (4) coordinate[c2,fill=colorV12V34]; \draw (4) coordinate[c1,fill=colorV13V24,label=95:$u$];
    \draw (5) coordinate[c2,fill=colorV13V24]; \draw (5) coordinate[c1,fill=darkgreen,label=0:$v_{1324}$];
    \draw (6) coordinate[c2,fill=colorV13V24]; \draw (6) coordinate[c1,fill=colorCurve24,label=290:$v_{24}$];
    \draw (7) coordinate[c2,fill=lightgray]; \draw (7) coordinate[c1,fill=colorV13V24,label=180:$v_{13}$];
    \draw (8) coordinate[c2,fill=lightgray]; \draw (8) coordinate[c1,fill=colorCurve24,label=30:$w_{24}$];
    \draw (9) coordinate[c2,fill=lightgray]; \draw (9) coordinate[c1,fill=lightgray,label={[label distance=1mm,align=center, font=\tiny]0:$\{v_{13}, v_{34}, w_{24}\}^\perp$ }];
    \draw (10) coordinate[c2,fill=lightgray]; \draw (10) coordinate[c1,fill=lightgray,label={[label distance=5mm,align=center, font=\tiny]0:$\{v_{13}, v_{14}, w_{23}\}^\perp$ }];
    \draw (11) coordinate[c2,fill=lightgray]; \draw (11) coordinate[c1,fill=lightgray,label={[label distance=5mm,align=center, font=\tiny]90:$\{v_{14}, v_{34}, w_{12}\}^\perp$ }];
    \draw (12) coordinate[c2,fill=lightgray]; \draw (12) coordinate[c1,fill=darkgreen,label=90:$w_{23}$];
    \draw (13) coordinate[c2,fill=lightgray]; \draw (13) coordinate[c1,fill=colorE4E1,label=285:$v_{14}$];
    \draw (14) coordinate[c2,fill=colorE4E1]; \draw (14) coordinate[c1,fill=darkgreen,label=180:$w_{14}$];
    \draw (15) coordinate[c2,fill=colorE4E1]; \draw (15) coordinate[c1,fill=colorCurve24,label=160:$e_4$];
    \draw (16) coordinate[c2,fill=colorE1E2]; \draw (16) coordinate[c1,fill=colorE4E1,label=0:$e_1$];
    \draw (17) coordinate[c2,fill=colorE1E2]; \draw (17) coordinate[c1,fill=colorCurve24,label=215:$e_2$];
    \draw (18) coordinate[c2,fill=colorE1E2]; \draw (18) coordinate[c1,fill=lightgray,label=180:$w_{12}$];

\end{tikzpicture}
\\
(a)&&(b)
\end{tabular}
\caption{
(a) Four-dimensional extension of the Harding--Salinas Schmeis hypergraph notation~\cite{harding2025remarksIJTP}.  Solid curves show the $13$ contexts of $\mathfrak{B}_{13}$: the $10$ imposed contexts and the three disjoint-support contexts $\mathcal{B}_{1234}$ (teal), $\mathcal{B}_{1324}$ (inner black circle), and $\mathcal{B}_{1423}$ (brown).  The dotted dark-green curve shows one of the four equality-induced contexts, $\mathcal{E}_3=\{w_{14},w_{23},v_{1234},v_{1324}\}$; the other three ($\mathcal{C}_0$, $\mathcal{E}_1$, and $\mathcal{E}_2$) are omitted for clarity and listed in the text.  (b) The Cabello $18$-ray, $9$-context KS configuration~\cite{cabello-96}.  Shared rays carry their gadget labels, and each of the three Cabello-only rays is labeled using three of the four shared rays lying on its two incident gray contexts, matching Table~\ref{tab:vector-equivalence} (which also lists Cabello's own labels~\cite{cabello:210401}); overall nonzero scalar factors are immaterial.  The six contexts shared with (a) keep their colors; the three Cabello-only contexts are gray.  The notation $\{v_i,v_j,v_k\}^{\perp}$ denotes the unique ray orthogonal to the span of three linearly independent rays.  The drawings record the stated incidence relations; no lattice-theoretic conclusion is inferred from the planar layout alone.
}
\label{fig:peres}
\end{figure*}

\begin{table*}[ht]
\caption{All $24$ rays of the Peres--Mermin eigensystem~\cite{peres111,mermin90b,peres-91,pavicic-2004ksafq,svozil-2024-convert-pra-externalfigures,2024-cabello-trandafir0}, comparing the $20$-ray gadget and the Cabello $18$-ray set~\cite{cabello-96,cabello:210401}; overall nonzero scalar factors are immaterial.  Fifteen rays are shared projectively, five belong only to the gadget, three only to Cabello's set, and $(1,-1,1,1)$ to neither.  Cabello's label $v^{\mathrm{C}}_{ij}$~\cite{cabello:210401} denotes the ray common to his contexts $i$ and $j$; the superscript distinguishes these labels from the gadget rays $v_{ij}$.  Where a ray is absent from one set, that column gives its construction $\{a,b,c\}^\perp$ from the other set.  Every construction in the Cabello column uses three mutually orthogonal rays (matching the contexts $\mathcal{B}_{\mathrm{comp}}$, $\mathcal{B}_{1423}$, $\mathcal{B}_{1324}$, $\mathcal{B}_{1234}$ of Fig.~\ref{fig:peres}(a), $\mathcal{C}_0$, and the $24$th Peres--Mermin context); in the gadget column no mutually orthogonal triple exists, and each of $v^{\mathrm{C}}_{34}$, $v^{\mathrm{C}}_{37}$, $v^{\mathrm{C}}_{47}$ uses three linearly independent rays from its two incident contexts in Fig.~\ref{fig:peres}(b).}
\label{tab:vector-equivalence}
\begin{ruledtabular}
\begin{tabular}{ccc}
Vector & Gadget Label / Construction & Cabello Label / Construction \\
\hline
$(1,1,1,1)$ & $u$ & $v^{\mathrm{C}}_{56}$ \\
$(1,0,0,0)$ & $e_1$ & $v^{\mathrm{C}}_{12}$ \\
$(0,1,0,0)$ & $e_2$ & $v^{\mathrm{C}}_{18}$ \\
$(0,0,1,0)$ & $e_3$ & $\{v^{\mathrm{C}}_{12}, v^{\mathrm{C}}_{18}, v^{\mathrm{C}}_{28}\}^\perp$ \\
$(0,0,0,1)$ & $e_4$ & $v^{\mathrm{C}}_{28}$ \\
\hline
$(0,0,1,-1)$ & $v_{12}$ & $v^{\mathrm{C}}_{16}$ \\
$(0,-1,0,1)$ & $v_{13}$ & $v^{\mathrm{C}}_{45}$ \\
$(0,1,-1,0)$ & $v_{14}$ & $v^{\mathrm{C}}_{23}$ \\
$(1,0,0,-1)$ & $v_{23}$ & $\{v^{\mathrm{C}}_{23}, v^{\mathrm{C}}_{29}, v^{\mathrm{C}}_{39}\}^\perp$ \\
$(-1,0,1,0)$ & $v_{24}$ & $v^{\mathrm{C}}_{58}$ \\
$(1,-1,0,0)$ & $v_{34}$ & $v^{\mathrm{C}}_{67}$ \\
\hline
$(0,0,1,1)$ & $w_{12}$ & $v^{\mathrm{C}}_{17}$ \\
$(0,1,0,1)$ & $w_{13}$ & $\{v^{\mathrm{C}}_{45}, v^{\mathrm{C}}_{58}, v^{\mathrm{C}}_{48}\}^\perp$ \\
$(0,1,1,0)$ & $w_{14}$ & $v^{\mathrm{C}}_{29}$ \\
$(1,0,0,1)$ & $w_{23}$ & $v^{\mathrm{C}}_{39}$ \\
$(1,0,1,0)$ & $w_{24}$ & $v^{\mathrm{C}}_{48}$ \\
$(1,1,0,0)$ & $w_{34}$ & $\{v^{\mathrm{C}}_{16}, v^{\mathrm{C}}_{67}, v^{\mathrm{C}}_{17}\}^\perp$ \\
\hline
$(1,1,-1,-1)$ & $v_{1234}$ & $v^{\mathrm{C}}_{69}$ \\
$(1,-1,1,-1)$ & $v_{1324}$ & $v^{\mathrm{C}}_{59}$ \\
$(1,-1,-1,1)$ & $v_{1423}$ & $\{v^{\mathrm{C}}_{56}, v^{\mathrm{C}}_{69}, v^{\mathrm{C}}_{59}\}^\perp$ \\
\hline
$(-1,1,1,1)$ & $\{v_{13}, v_{14}, w_{23}\}^\perp$ & $v^{\mathrm{C}}_{34}$ \\
$(1,1,1,-1)$ & $\{v_{14}, v_{34}, w_{12}\}^\perp$ & $v^{\mathrm{C}}_{37}$ \\
$(1,1,-1,1)$ & $\{v_{13}, v_{34}, w_{24}\}^\perp$ & $v^{\mathrm{C}}_{47}$ \\
\hline
$(1,-1,1,1)$ & $\{v_{12}, v_{23}, w_{34}\}^\perp$ & $\{v^{\mathrm{C}}_{34}, v^{\mathrm{C}}_{37}, v^{\mathrm{C}}_{47}\}^\perp$ \\
\end{tabular}
\end{ruledtabular}
\end{table*}

\subsubsection{Concrete \texorpdfstring{$D=4$}{D=4} example: \texorpdfstring{$u=(1,1,1,1)$}{u=(1,1,1,1)}}

Take the center vector $u=(1,1,1,1)$ and the rim vectors
\begin{align*}
e_1&=(1,0,0,0),& e_2 =(0,1,0,0),\\
e_3&=(0,0,1,0),& e_4 =(0,0,0,1).
\end{align*}
The pair minors and their complementary vectors become:
\begin{align*}
v_{12}&=(0,0,1,-1), & w_{12}&=(0,0,1,1),\\
v_{13}&=(0,-1,0,1), & w_{13}&=(0,1,0,1),\\
v_{14}&=(0,1,-1,0), & w_{14}&=(0,1,1,0),\\
v_{23}&=(1,0,0,-1), & w_{23}&=(1,0,0,1),\\
v_{24}&=(-1,0,1,0), & w_{24}&=(1,0,1,0),\\
v_{34}&=(1,-1,0,0), & w_{34}&=(1,1,0,0).
\end{align*}
Each pair block
\[
\mathcal{B}_{ij}=\{e_i, e_j, v_{ij}, w_{ij}\}
\]
forms an ONB after normalizing $v_{ij}$ and $w_{ij}$.

The connector vectors are
\begin{align*}
v_{1234}&=(1,1,-1,-1),\\
v_{1324}& =(1,-1,1,-1),\\
v_{1423}&=(1,-1,-1,1),
\end{align*}
and they are all orthogonal to $u=(1,1,1,1)$ by inspection, so the connector blocks $\mathcal{C}_{(a)}$, $\mathcal{C}_{(b)}$, and $\mathcal{C}_{(c)}$ form orthogonal bases after normalization.

Together with $\mathcal{B}_{\mathrm{comp}}$, the six pair contexts and three connector contexts form the 10-context scaffold $\mathfrak{B}_{10}$ defined below.  The forcing in Proposition~\ref{prop:D4forcing} rests on the prescribed coordinates of $v_{1234},v_{1324},v_{1423}$; the incidences of $\mathfrak{B}_{10}$ alone do not force it.  Indeed, for every $u$ with nonzero coordinates the rays $v_{ij}$ and $w_{ij}$ are fixed by their contexts, while $v_{1234}$ occurs only in $\mathcal{C}_{(a)}$ and may be taken as the ray orthogonal to $u$, $v_{12}$, and $v_{34}$,
\[
\tilde v_{1234}=\bigl(s_{34}x_1,\,s_{34}x_2,\,-s_{12}x_3,\,-s_{12}x_4\bigr)
\]
with $s_{kl}=|x_k|^2+|x_l|^2$, and analogously $\tilde v_{1324}$ and $\tilde v_{1423}$ from $\mathcal{C}_{(b)}$ and $\mathcal{C}_{(c)}$.  Hence $\mathfrak{B}_{10}$ can be realized with a center vector that is not unbiased.  The equal-modulus condition also creates $\mathcal{C}_0$ and the three contexts $\mathcal{E}_1,\mathcal{E}_2,\mathcal{E}_3$ listed above.  For example, $\langle w_{14},v_{1234}\rangle=|x_2|^2-|x_3|^2=0$; $\mathcal{E}_3$ is the dotted dark-green curve in Fig.~\ref{fig:peres}(a).  Conversely, the complete $17$-context hypergraph forces maximal unbiasedness by its incidences alone.  The contexts $\mathcal{E}_3$ and $\mathcal{E}_1$ require $\langle w_{14},\tilde v_{1234}\rangle=\langle w_{14},\tilde v_{1324}\rangle=\langle w_{12},\tilde v_{1324}\rangle=0$, that is,
\[
|x_2x_4|=|x_1x_3|,\quad
|x_3x_4|=|x_1x_2|,\quad
|x_2x_3|=|x_1x_4|,
\]
whose only positive solution is $|x_1|=|x_2|=|x_3|=|x_4|$.

\subsubsection{Mutual reconstruction inside the Peres--Mermin eigensystem}

Both ray sets are subsets of the $24$-ray Peres--Mermin eigensystem~\cite{pavicic-2004ksafq,svozil-2024-convert-pra-externalfigures,2024-cabello-trandafir0}.  They share $15$ rays projectively, leaving five gadget-only rays, three Cabello-only rays, and one remaining Peres--Mermin ray, $(1,-1,1,1)$, that lies in neither set (cf.\ Ref.~\cite[Table~III]{2024-cabello-trandafir0}).  Table~\ref{tab:vector-equivalence} lists all $24$ rays together with rank-three orthogonal-complement constructions for every ray missing from either set.  Under the Hilbert-space operation ``take the unique ray orthogonal to three linearly independent rays,'' each ray set generates the other (and the full $24$-ray eigensystem); this is the precise sense in which we call them informationally equivalent.  A sharper asymmetry appears if one restricts to three \emph{mutually orthogonal} rays (basis completion): the six incomplete orthogonal triples of Cabello's $18$-ray set immediately produce all five gadget-only rays and $(1,-1,1,1)$, whereas the $20$-ray gadget is already closed under basis completion (all $68=17\times 4$ of its orthogonal triples extend inside it) and therefore requires nonorthogonal independent triples to generate $v^{\mathrm{C}}_{34}$, $v^{\mathrm{C}}_{37}$, $v^{\mathrm{C}}_{47}$, and $(1,-1,1,1)$.

Informational equivalence does not imply equality of context hypergraphs.  The complete $20$-ray orthogonality hypergraph has $17$ contexts and $16$ separating two-valued states.  The Cabello $18$-ray hypergraph has nine contexts, which are all the orthogonal quadruples among its $18$ rays, and it has no two-valued state; its parity argument makes it a KS configuration~\cite{cabello-96}.  Both hypergraphs are thus complete on their own ray sets.  The contrast is therefore not a consequence of ``more contexts allowing loopholes''---adding constraints cannot create two-valued states---but of the two constructions selecting different rays and different contexts from their common $24$-ray parent.

\subsubsection{Context completion and two-valued states}

The context choice within the $20$-ray set can itself be tracked explicitly.  Let
\[
\mathfrak{B}_{10}=\{\mathcal{B}_{\mathrm{comp}},\mathcal{B}_{ij}\ (1\leq i<j\leq4),
\mathcal{C}_{(a)},\mathcal{C}_{(b)},\mathcal{C}_{(c)}\}
\]
be the 10 imposed contexts, and let
\[
\mathfrak{B}_{13}=\mathfrak{B}_{10}\cup
\{\mathcal{B}_{1234},\mathcal{B}_{1324},\mathcal{B}_{1423}\}.
\]
Exact enumeration shows that both hypergraphs have the same $36$ two-valued states, and those states are separating and unital.  The three disjoint-support contexts therefore record exclusions already obeyed by every state of $\mathfrak{B}_{10}$; they do not remove an assignment.  Adding the four equality-induced contexts gives the complete set
\[
\mathfrak{B}_{17}=\mathfrak{B}_{13}\cup\{\mathcal{C}_0,\mathcal{E}_1,\mathcal{E}_2,\mathcal{E}_3\},
\]
which has $16$ separating and unital two-valued states.  This nested comparison makes the monotonicity transparent: adding contexts can preserve or reduce the state set, but cannot enlarge it.  It also avoids attributing KS contextuality to an incomplete incidence structure merely because some available contexts were omitted.

The apparent asymmetry of Fig.~\ref{fig:peres}(a) is graphical: among the four equality-induced contexts $\mathcal{C}_0,\mathcal{E}_1,\mathcal{E}_2,\mathcal{E}_3$, only $\mathcal{E}_3$ is drawn, whereas the complete $17$-context hypergraph $\mathfrak{B}_{17}$ is symmetric under all coordinate permutations.  Both it and the Cabello $18$-ray set sit inside the $24$-ray, $24$-context Peres configuration~\cite{Pavicic-2005,pavicic-2005csvcorri}; neither subset needs to
inherit all symmetries of the parent.

\subsection{A \texorpdfstring{$D=5$}{D=5} scaffold and forcing constraint}

Let $\varepsilon_{abcde}$ be a totally antisymmetric tensor
with $\varepsilon_{12345}=1$.  For each ordered triple $(i,j,k)$ of distinct indices (say $1\le i<j<k\le 5$) define
\[
(v_{ijk})_m=\sum_{\ell=1}^5\varepsilon_{mijk\ell}\,\overline{x_\ell}.
\]
The vector $v_{ijk}$ vanishes in coordinates $i,j,k$ and satisfies $\langle v_{ijk},u\rangle=0$ by antisymmetry.  Choose $w_{ijk}$ on the complementary two coordinates with $v_{ijk}\perp w_{ijk}$; then
\[
\mathcal{B}_{ijk}=\{e_i,e_j,e_k,v_{ijk},w_{ijk}\}
\]
is a scaffold context.

The four connector rays are represented by
\begin{align*}
g_1&=(x_1,0,0,0,-x_5),&
g_2&=(0,x_2,0,0,-x_5),\\
g_3&=(0,0,x_3,0,-x_5),&
g_4&=(0,0,0,x_4,-x_5).
\end{align*}
Whenever $u\perp g_i$, three rays spanning $\operatorname{span}\{u,g_i\}^{\perp}$ complete the connector context $\mathcal{C}_i=\{u,g_i,h^{(i)}_1,h^{(i)}_2,h^{(i)}_3\}$.
Unlike in $D=4$, the scaffold contexts $\mathcal{B}_{ijk}$ are not needed for Proposition~\ref{prop:D5forcing}; they are the $D=5$ analogue of the pair blocks $\mathcal{B}_{ij}$.  They can supply completion rays: for $m\notin\{i,5\}$, the ray $v_{abc}$ with $\{a,b,c\}=\{i,m,5\}$ vanishes on the support $\{i,5\}$ of $g_i$ and is orthogonal to $u$, so it may serve as $h^{(i)}_1$.  As in $D=4$, the forcing uses the prescribed coordinates of the connector rays $g_i$.

\begin{proposition}[$D=5$ forcing constraint]
\label{prop:D5forcing}
With the connector coordinates $g_1,\ldots,g_4$ given above, membership of $u$ and $g_i$ in the four connector contexts $\mathcal{C}_i$ forces a normalized center vector $u$ to be maximally unbiased relative to the computational basis.
\end{proposition}

\begin{proof}
For $i=1,2,3,4$,
\[
\langle u,g_i\rangle=|x_i|^2-|x_5|^2.
\]
Thus the four orthogonality conditions imply \ensuremath{|x_1|=\cdots=|x_5|};
normalization makes every squared modulus $1/5$.
\end{proof}

\section{Summary and Conclusion}


The qutrit construction yields $165$ projectively distinct rays and exactly $130$ contexts.  Correct provenance accounting is symmetric: the three $69$-ray, $50$-context KS lineages share $21$ rays and $10$ contexts, and each adds $48$ exclusive rays and $40$ lineage-specific contexts.  This places the triple of $25$-ray full-context Yu--Oh completions, the Harding--Salinas Schmeis and Cabello descriptions, and the $69$-$50$ hypergraph of Pavi\v{c}i\'{c} and Megill~\cite{pavicic-megill-2022,Pavicic_2023} in a common MUB-based framework without assigning shared rays to an arbitrary first lineage.

In $D=4$ and $D=5$, the forcing mechanism is explicit at the level of constraints: for the prescribed connector coordinates, connector orthogonalities yield linear equations in the squared coordinate moduli and hence maximal unbiasedness.  In $D=4$, the complete $17$-context hypergraph forces the same conclusion by its incidences alone.  For the concrete $D=4$ realization, the complete $20$-ray orthogonality hypergraph has $17$ contexts and $16$ separating two-valued states.  Its ray set is mutually reconstructible with the Cabello $18$-ray set inside the $24$-ray Peres--Mermin eigensystem, but the Cabello nine-context hypergraph is uncolorable.  Ray-set closure and context incidence are therefore distinct pieces of data.

The main methodological conclusion is that contextuality claims should specify the incidence hypergraph being tested and, when completeness is intended, enumerate every context supported by the chosen rays.  Restricting attention to intertwining rays or to a selected subset of available contexts may define a legitimate reduced scenario, but its two-valued-state properties should not be transferred to a different completion without a separate check.

\begin{acknowledgments}
We are grateful to Josef Tkadlec for providing a {\em Pascal} program that computes and analyses the set of two-valued states of collections of contexts.
We are also grateful to Ad\'{a}n Cabello and Mladen Pavi{\v{c}}i{\'{c}} for valuable communications.

OpenAI Codex (GPT-5) was used to assist with literature organization, \LaTeX{}
editing, and checks of derivations.  The authors
specified the scientific
direction and prompts; model output retained in the manuscript was checked
against the cited sources and explicit calculations.  The authors accept
full
responsibility for the final text.

This research was funded in part by the Austrian
Science Fund (FWF), Grant DOI \href{https://doi.org/10.55776/PIN5424624}{10.55776/PIN5424624},
and the Czech Science
Foundation (GA\v{C}R), Grant No.~25-20013L.

The authors declare no conflict of interest.
\end{acknowledgments}

\bibliography{svozil}

\appendix

\section{Mutually unbiased bases in \texorpdfstring{$\mathbb{C}^3$}{C\textasciicircum 3}}
\label{appendix:MUBsD3}

For introductions to MUBs, see Refs.~\cite{Schwinger.60,WooFie,Brierley2009,springerlink:10.1007/978-3-540-24633-6_10,durt,mcnulty2024mutually}.  Write
\[
\omega=\mathrm{e}^{2\pi\ii{}/3},\qquad
\omega^3=1,\qquad 1+\omega+\omega^2=0,
\qquad \overline{\omega}=\omega^2.
\]
In $D=3$, a complete MUB family contains four bases~\cite{ivanovic-81,ivanovic-1997}.  Omitting the common normalization factor $1/\sqrt{3}$ from the last three bases, we use
\begin{align}
\mathcal{B}_0&=\{(1,0,0),(0,1,0),(0,0,1)\}, \label{eq:B0_3}\\
\mathcal{B}_1&=\{(1,1,1),(1,\omega,\omega^2),(1,\omega^2,\omega)\}, \label{eq:B1_3}\\
\mathcal{B}_2&=\{(1,\omega,\omega),(1,\omega^2,1),(1,1,\omega^2)\}, \label{eq:B2_3}\\
\mathcal{B}_3&=\{(1,\omega^2,\omega^2),(1,\omega,1),(1,1,\omega)\}. \label{eq:B3_3}
\end{align}
Here $\mathcal{B}_0$ is computational and $\mathcal{B}_1$ is the Fourier basis.  Direct use of $1+\omega+\omega^2=0$ verifies orthogonality within each basis and $|\langle b,b'\rangle|^2=1/3$ between normalized vectors from different bases.  The ordering of $\mathcal{B}_2$ and $\mathcal{B}_3$ agrees with 
the seed-vector groups assigned to columns 1--3, 4--6, and 7--9 of Table~\ref{tab:vector_generation}.

For the higher-dimensional argument, no catalogue of a~complete MUB family is required: Propositions~\ref{prop:D4forcing} and~\ref{prop:D5forcing} use only the criterion \ensuremath{|x_1|=\cdots=|x_D|} for one vector to be maximally unbiased relative to the computational basis.

\end{document}